\documentclass[twocolumn]{aa} % for the letters 
\usepackage[]{natbib}
\usepackage{graphicx,color}
\usepackage{txfonts}
\usepackage[colorlinks=true, citecolor=blue]{hyperref}
\usepackage[table]{xcolor}
\usepackage{multirow} %multiple row in tables
\usepackage{boldline} %bold lines in tables
\usepackage{adjustbox}
\usepackage{float}
\usepackage{booktabs}

\newcommand{\orcid}[1]{\href{https://orcid.org/#1}{\includegraphics[width=10pt]{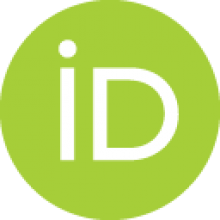}}}

\begin{document} 

\title{
Multi-Wavelength characterization of VVVX open clusters}

\author{
C.O. Obasi\inst{1}\thanks{Casmir Obasi: casmir.obasi@ucn.cl}
\and E. R. Garro\inst{2}
\and J.G. Fern\'andez-Trincado\inst{1}\thanks{Jos\'e G. Fern\'andez-Trincado: jose.fernandez@ucn.cl}\orcid{0000-0003-3526-5052}
\and D. Minniti\inst{3,4}
\and M. G\'omez\inst{3}
\and M. C. Parisi\inst{5,6}
\and M. Ortigoza-Urdaneta\inst{7}
}

\authorrunning{C.O.Obasi et.:} 
	
\institute{
Instituto de Astronom\'ia, Universidad Cat\'olica del Norte, Av. Angamos 0610, Antofagasta, Chile.
\and ESO - European Southern Observatory, Alonso de Cordova 3107, Vitacura, Santiago, Chile
\and Instituto de Astrofísica, Depto. de Ciencias Fisicas, Facultad de Ciencias Exactas, Universidad Andr\'es Bello, Av. Fern\'andez Concha 700, Las Condes, Santiago, Chile
\and Vatican Observatory, V00120 Vatican City State, Italy
    \and Observatorio Astronómico, Universidad Nacional de Córdoba, Laprida 854, X5000BGR Córdoba, Argentina\and Instituto de Astronomía Teórica y Experimental (CONICET-UNC), Laprida 854, X5000BGR Córdoba, Argentina
    \and Departamento de Matemática, Facultad de Ingeniería, Universidad de Atacama, Copiapó, Chile}

	\date{Received ...; Accepted ...}
	\titlerunning{Confirmation and characterization of nine open clusters}
	
	% \abstract{}{}{}{}{} 
	% 5 {} token are mandatory
	
	% \abstract{}{}{}{}{} 
% 5 {} token are mandatory
 
  \abstract
   {With the rise of large surveys across wavelengths, both supervised and unsupervised machine learning algorithms have increasingly aided in detecting large samples of old open clusters in high-extinction regions of the Milky Way bulge and disk.}
   {Our primary goal is to confirm or discard automatically detected open clusters from poorly studied, heavily contaminated regions of the Milky Way. Cleaning these samples is critical for reconstructing the Galactic disk's star formation history and understanding the thin and thick disk formation model.}
   {We used data from the VISTA Variables in the Via Láctea extended Survey (VVVX), Two Micron All-Sky Survey (2MASS), and \textit{Gaia} Data Release 3 (\textit{Gaia} DR3) to confirm and characterize nine open cluster candidates: BH118, BH 144, Schuster-MWSC 1756, Saurer 3, FSR 1521, Saurer 2, Haffner 10-FSR 1231, Juchert 12, and Pismis 3. We constructed density maps and vector-proper motion diagrams to analyse the targets. Additionally, we performed photometric analysis to derive their main physical parameters.
   }
   {We examined cluster images in 2MASS, Wide-field Infrared Survey Explorer (WISE), and  Dark Energy Camera Plane Survey (DECaPS), identifying star clusters through an over-density of stars. This was confirmed with a VVVX photometry density map and validated using Gaussian Kernel Density Estimation. Using \textit{Gaia} proper motion data, we refined cluster memberships and decontaminated the data to build the final cluster catalogue with high-likely star cluster members.  We derived the following parameters: extinction values ranging from A$_{Ks}$=0.07$\pm$0.03  to 0.50$\pm$0.04 ; colour excess values from  E(J-K$_s$)= 0.16$\pm$0.03  to 0.60$\pm$0.03 ; distances from D = 2.19$\pm$0.06 kpc to 8.94$\pm$0.06 kpc; Galactocentric distances from R$_G$= 7.82 kpc to 15.08 kpc; vertical distance component values from Z = -0.09 kpc to 0.34 kpc; and tangential velocities from V$_T$= 30.59 km/s to  245.42 km/s.  We also computed age and metallicity by fitting Parsec isochrones, finding ages ranging from t= 20 Myr to 5 Gyr and metallicity from [Fe/H] = -0.5 to 0.5. Structural parameters include core radii from r$_c$ = 0.71' to 5.21', tidal radii from r$_t$ = 3.4' to 12.0', and concentration indices from c = 0.36 to 0.83.
}
   {we photometrically confirmed the open cluster nature for 9 targets in our compilation, updating their main physical parameters.}

	\maketitle

%%%%% INTRODUCTION %%%%%%%
\section{Introduction}
\label{section1}
In the last three decades, the interest  in
the studies of intermediate-age (ages between 100 Myr and 1 Gyr) and old Galactic Open Clusters (OCs; those older than 1 Gyr) have increased significantly
\citep[see e.g;][]{montgomery1994reddening,phelps1994janes,friel1995old,tosi1998old,frinchaboy2004star,tosi2007old,friel2010abundances,angelo2019investigating,tarricq2022structural}. The reason is that understanding their properties like ages, metallicities, and kinematics is fundamental to better discern the structure and formation
history of the thin and thick disk \citep{carraro1994galactic}.

 In the era of extensive surveys, such as the Two Micron All-Sky Survey (2MASS) \citep{skrutskie2006two} and  \textit{Gaia} \citep{gaia2016gaia}, there has been an explosion in the discovery of many new OCs. \cite{kronberger2006new} used Sloan Digital Sky Survey  \citep[SDSS;][]{york2000sloan} and 2MASS images of selected Milky Way regions to discover 66 stellar groupings. The morphologies, colour-magnitude diagrams, and stellar density distributions of these objects suggest they are potential OCs and none of them have previously been listed in any catalogue. In these studies, they provided extensive descriptions for 24 candidates considered the most likely OCs based on 2MASS photometry. Of these 24, only 9 objects had fundamental parameters determined by fitting the colour-magnitude diagrams with solar metallicity \textit{Padova} isochrones \citep{marigo2017new}. An additional 10 cluster candidates had distance modulus and reddening derived from K$_{s}$ magnitudes and colour indices of helium-burning red clump stars.

The catalogue by \cite{kharchenko2013global} identified 3,006 real objects from an initial list of 3,784 targets obtained from various sources. While most of these objects are classified as OCs, the catalogue also includes stellar associations and globular clusters. It provides essential cluster parameters such as central position, apparent size, proper motion (PM), distance, colour excess, and age. However, 778 of the 3,006 objects were not confirmed.

In another study, \cite{cantat2020painting} identified 2,000 clusters using \textit{Gaia} DR2 astrometry. Using deep \textit{Gaia} photometry down to G=18, they measured the distance, age, and reddening of these clusters. An artificial neural network, trained on previously documented objects with well-defined parameters, was employed to estimate the parameters of these new clusters. Out of the 2,000 clusters analyzed, reliable parameters were obtained for 1,867.

The VVVX survey has played a central role in confirming many new globular and OCs within the Galactic bulge and central region, where spatial extinction is high \citep{saito2024vista,obasi2023globular,obasi2024searching,garro2022unveiling,garro2022new,garro2024vvvx}.
Recently, we utilized VVVX data to discover and characterize a new open cluster, Garro 3 \citep{garro2024vvvx}. 

Recent OC studies have also relied heavily on the \textit{Gaia} survey for example, 628 OCs were discovered using the open cluster finder algorithm with  \textit{Gaia} EDR3 data by \cite{castro2022hunting}, and 704 new potential clusters were found with the sample-based clustering search method using  \textit{Gaia} EDR3 data by \cite{hao2022newly}. A machine-learning approach was developed using  \textit{Gaia}, with 23 new OCs in the Tycho- \textit{Gaia} Astrometric Solution dataset \citep{castro2018new}. \cite{cantat2019gaia} found 41 new OCs towards the Perseus Arm. Closer to this, \cite{castro2019hunting} detected 53 new OCs towards the Galactic anti-center. These are just a few of the newly discovered OCs, with many more not mentioned here. However, many detected but unconfirmed clusters still lack characterization of their physical parameters.

\begin{figure*}[t]
\begin{center}
\includegraphics[height=4.5cm]{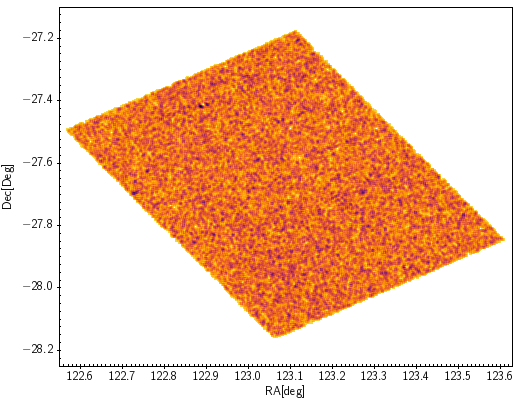}
\includegraphics[height=4.5cm]{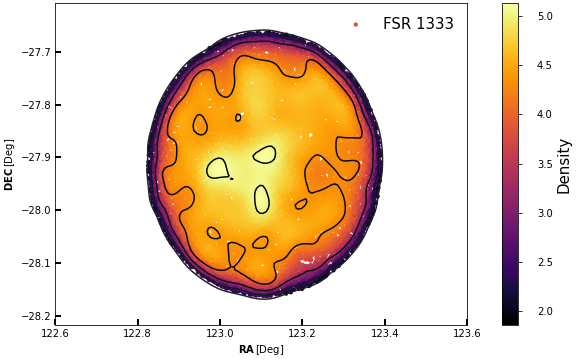}\\
\includegraphics[height=4.5cm]{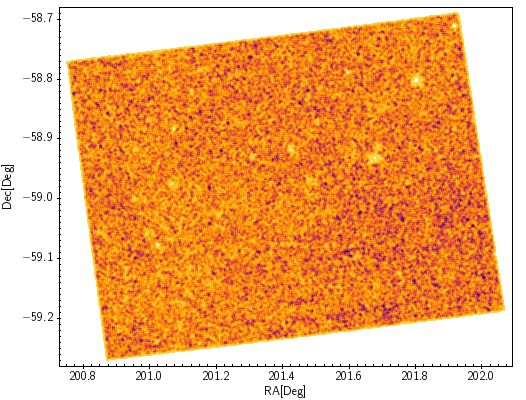}
\includegraphics[height=4.5cm]{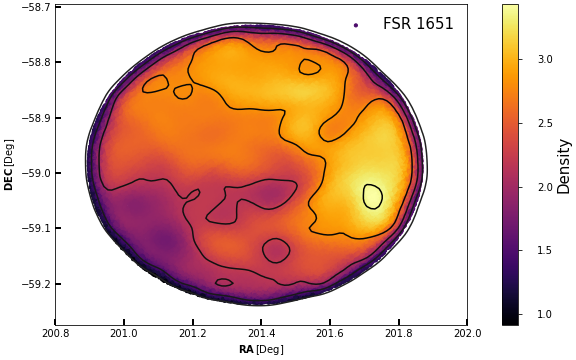}

\caption{VVVX density maps and Gaussian kernel density estimations for the clusters FSR 1333 and FSR 1651 are shown in the upper and lower panels left-right, respectively.
}
\label{Figure1A}
\end{center}
\end{figure*}

Consequently, many new findings are considered potential OC candidates, and some are false positives when better observations are made. For example, FSR 1333, with J2000 coordinates $\alpha$= 123.108, $\delta$= -27.9130, and FSR 1651, with J2000 coordinates $\alpha$= 201.382, $\delta$= -58.9850, were both classified as open clusters by \cite{buckner2013properties}. However, our current analysis using VVVX data revealed no over-density in the VVVX photometry density map and Gaussian Kernel density estimation (KDE), as shown in Figure \ref{Figure1A}. This suggests that both clusters might be false positive detections.

Therefore, it remains essential to confirm so many of these new detections and measure their physical parameters. In this context, one of the primary objectives of the VVV Team is to confirm new star clusters within the VVV, \citep{minniti2010vista,saito2012vvv} and its extension VVVX \citep{saito2024vista} based on observations at the VISTA 4m telescope. In this task, the importance of the VVV lies in its operating wavelength at the near-infrared (NIR), which has proved to be much more effective in the highly crowded central region of the Galactic bulge where high spatial extinction exists.

 In this paper, we use a combination of VVVX, 2MASS, and optical \textit{Gaia} DR3 \citep{brown2021gaia} data listed in Table \ref{table1} to confirm and characterize the physical nature of nine selected OCs toward the Galactic bulge. Additionally, we provide the first physical measurements and characterization of Jurchert12. Section \ref{section2} describes the dataset sources used in this work. Section \ref{section3} outlines the methodology for the analysis. In Section \ref{section4}, we present our derived physical parameters and a summary of the results. Section \ref{section5} presents the results, while Sections \ref{section6} and \ref{section7} 
 present the discussion and conclusions.

%%%%% DATA %%%%%%%
\section{Datasets}
\label{section2}

In this work, VVVX data in combination with 2MASS and  \textit{Gaia} DR3 was used to characterize nine OC candidates: BH118 , BH144,  Schuster-MWSC1756, Saurer3, FSR1521, Saurer2, Haffner10-FSR1231, Juchert12 and Pismis3 listed in the catalogues of \cite{kronberger2006new,kharchenko2013global} and \cite{cantat2020painting}. These clusters had not been properly characterized before. Our datasets comprised both NIR and optical data, and we used a matching radius of 0.5". The NIR data came from the VVVX and the 2MASS \citep{saito2024vista,skrutskie2006two}. The VVVX is a public ESO survey, conducted with the VISTA Infrared Camera (VIRCAM) on the 4.1 m wide-field Visible and Infrared Survey Telescope for Astronomy \citep[VISTA;][]{emerson2010visible}. 

The point spread function (PSF) photometry was extracted as described by \cite{alonso2018milky}. Astrometry was calibrated to the  \textit{Gaia} Data Release 3 \citep[DR3;][]{brown2021gaia}  reference system, while photometry was calibrated to the VISTA magnitude system \citep{gonzalez2018vista} against the 2MASS using a globally optimized model of frame-by-frame zero points plus an illumination correction. We used NIR 2MASS data to reach magnitudes brighter than Ks = 11, which are saturated in the VVVX images. Although we treated the VVVX and 2MASS data separately, we scaled them to the same photometric system \citep{gonzalez2018vista}, applying correction offsets of $\Delta$Ks < 0.009  and $\Delta$J < 0.050 .

Previous studies on these clusters by \cite{kronberger2006new,kharchenko2013global}, and \cite{cantat2020painting} have produced conflicting results, with no generally agreed-upon properties. The inspection of the VVVX tiles that contain these clusters revealed a clear and well-separated overdensity of red giant stars at their equatorial coordinates, listed in Table \ref{table1}.
\begin{table*}[]
\caption{Cluster name, coordinates (RA and Dec), Galactic coordinates ($l$ and b), VVVX tile corresponding to each cluster, and literature references.}
\label{table1}
\resizebox{\textwidth}{!}{%
\begin{tabular}{@{}lllllll@{}}
\toprule
Name              & Ra (h:m:s)           & Dec (d:m:s)         & $l$        & $b$       & VVVX-Tile Number & Ref            \\ \midrule
BH118             & 11:23:03.600   & $-$58:32:56.40  & 291.6041 & 2.3655  & e809        & (a)     \\
BH\_144           & 13:15:20.424 & $-$65:55:23.88 & 305.3625 & $-$3.1619 & e736-735    & (b) \\
Schuster-MWSC1756 & 10:04:36.192 & $-$55:51:24.84 & 281.0043 & $-$0.2568 & e1081       & (b) \\
Saurer3           & 10:41:30.000     & $-$55:18:00.00    & 285.1036 & 3.0091  & e805        & (b) \\
FSR1521           & 09:55:22.992 & $-$56:36:05.76 & 280.4353 & $-$1.6251 & e1035       & (b) \\
Saurer2           & 08:25:28.800   & $-$39:37:48.00    & 257.9931 & $-$0.9970 & e1065       & (b)              \\
Haffner10-FSR1231 & 07:28:34.800   & $-$15:21:54.00    & 230.7990 & 1.01100   & e1091       & (b) \\
Juchert12&07:20:56.712&$-$22:52:00.12&236.5614& $-$4.1232&e605&(c)\\
Pismis3&08:31:15.600&$-$38:39:28.80&257.8590 & 0.4803&e1110&(b)\\
\bottomrule

\end{tabular}%

}
\begin{flushleft}
\textbf{References:} (a) \citep{cantat2020painting}, (b) \citep{kharchenko2013global}, (c) \citep{kronberger2006new}
\end{flushleft}

\end{table*}
This confirms their status as real star clusters. Additionally, we visually inspected all the cluster overdensities in the 2MASS, WISE, and DECaPS images to ensure that these features were not artefacts and remained consistent under small variations of parameters (i.e., observing the targets in different wavelengths).

 \begin{figure*}[t]
\begin{center}
\includegraphics[height = 4.5 cm]{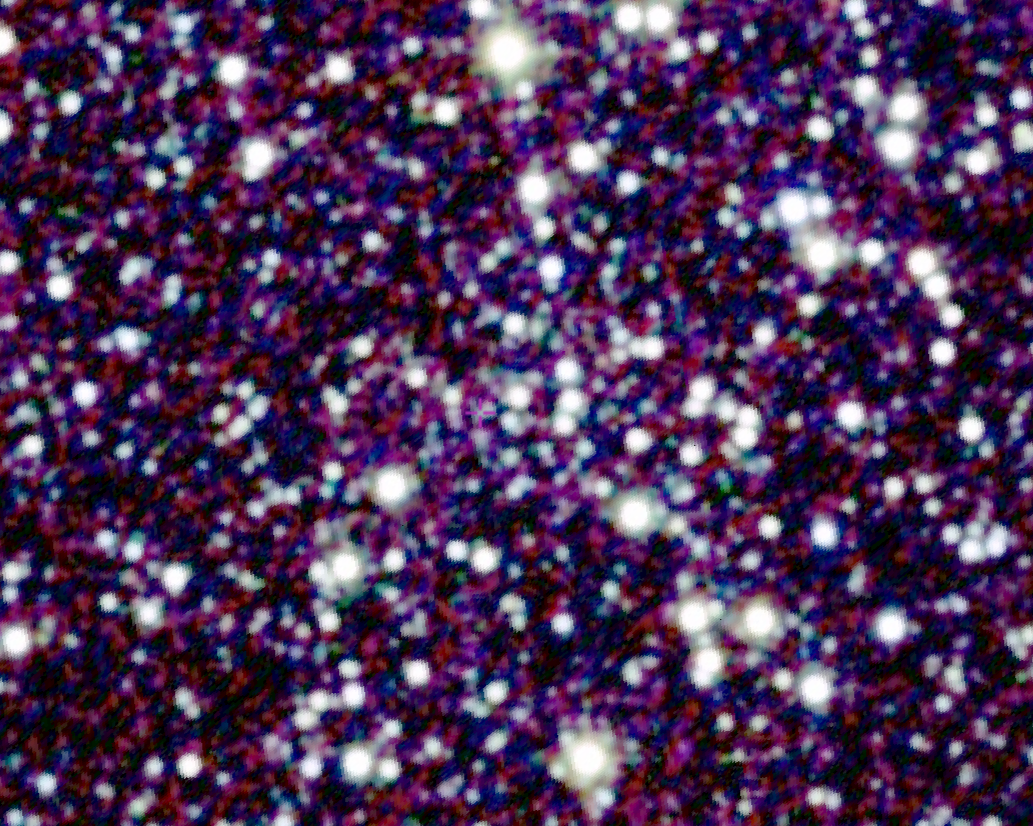}
\includegraphics[height = 4.5cm]{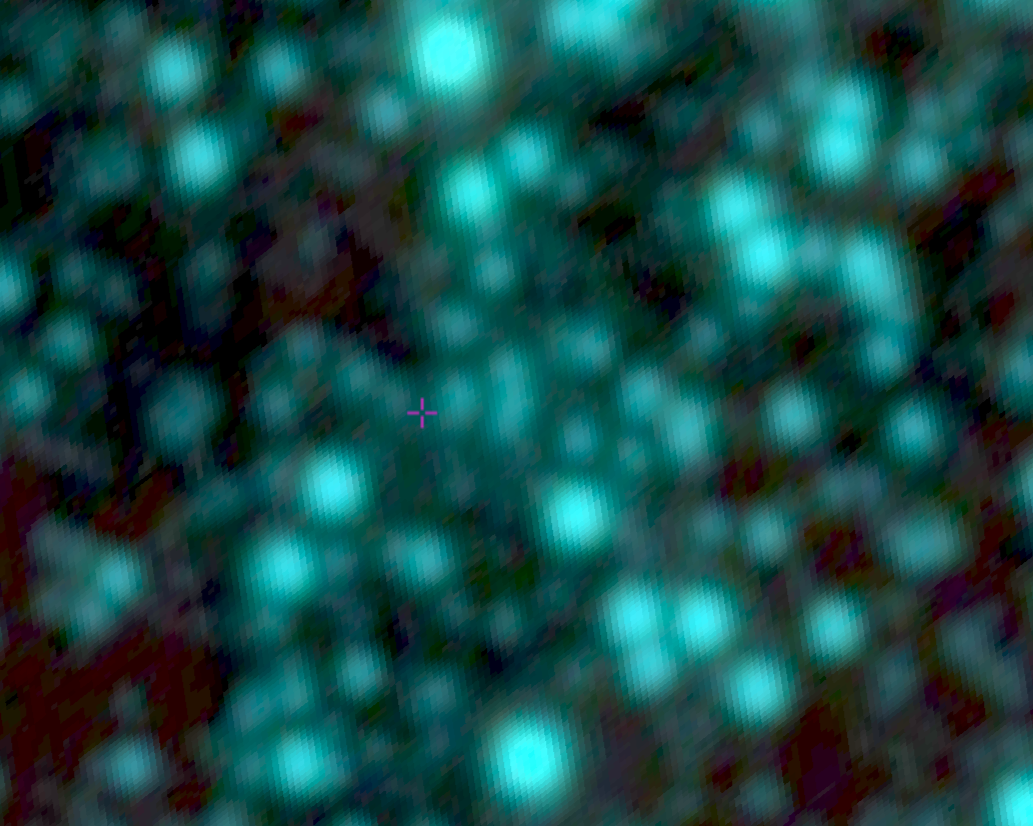}
\includegraphics[height = 4.5cm]{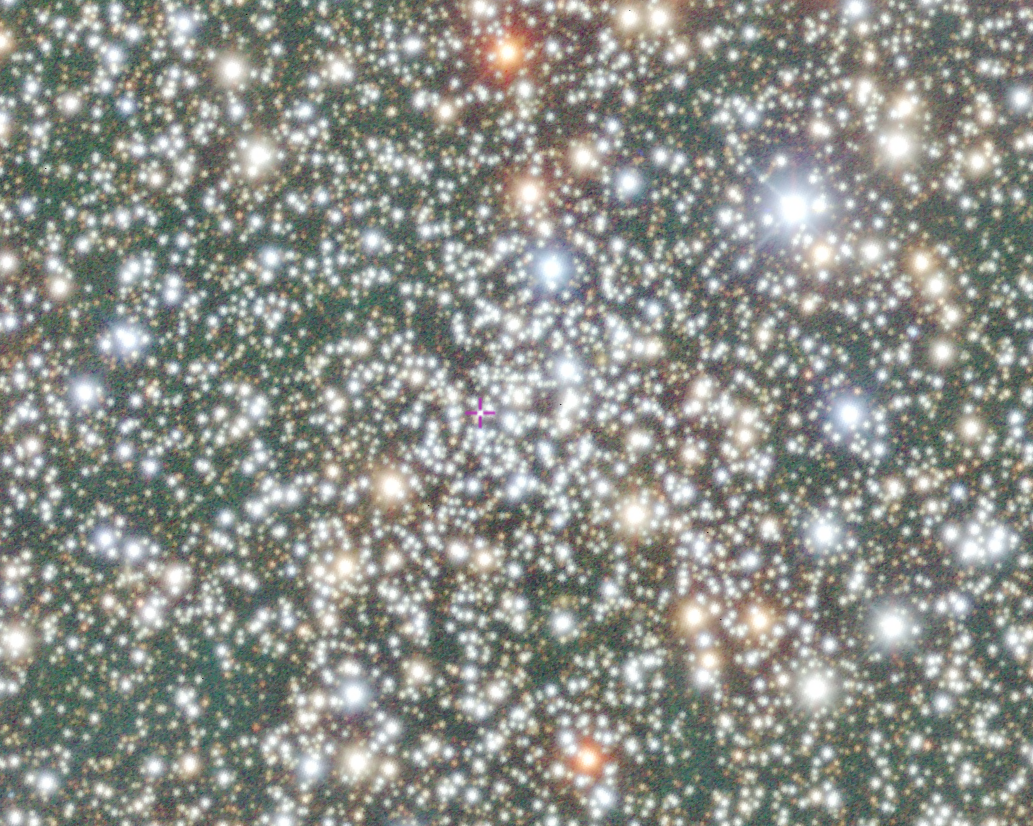}\\
\includegraphics[height = 5cm]{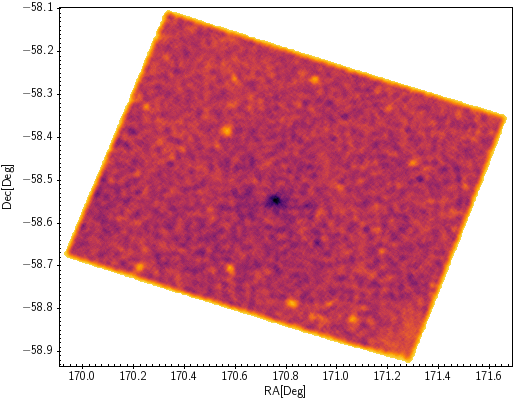}
\includegraphics[height = 5cm]{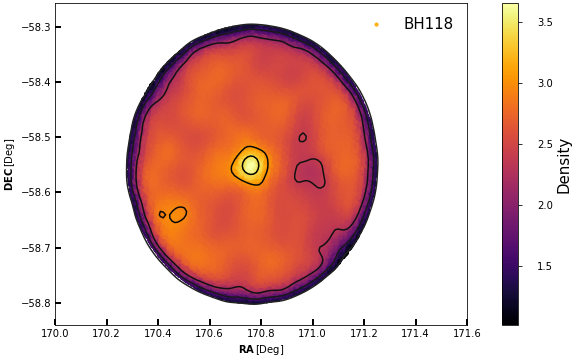}\\
\includegraphics[height = 4.5cm]{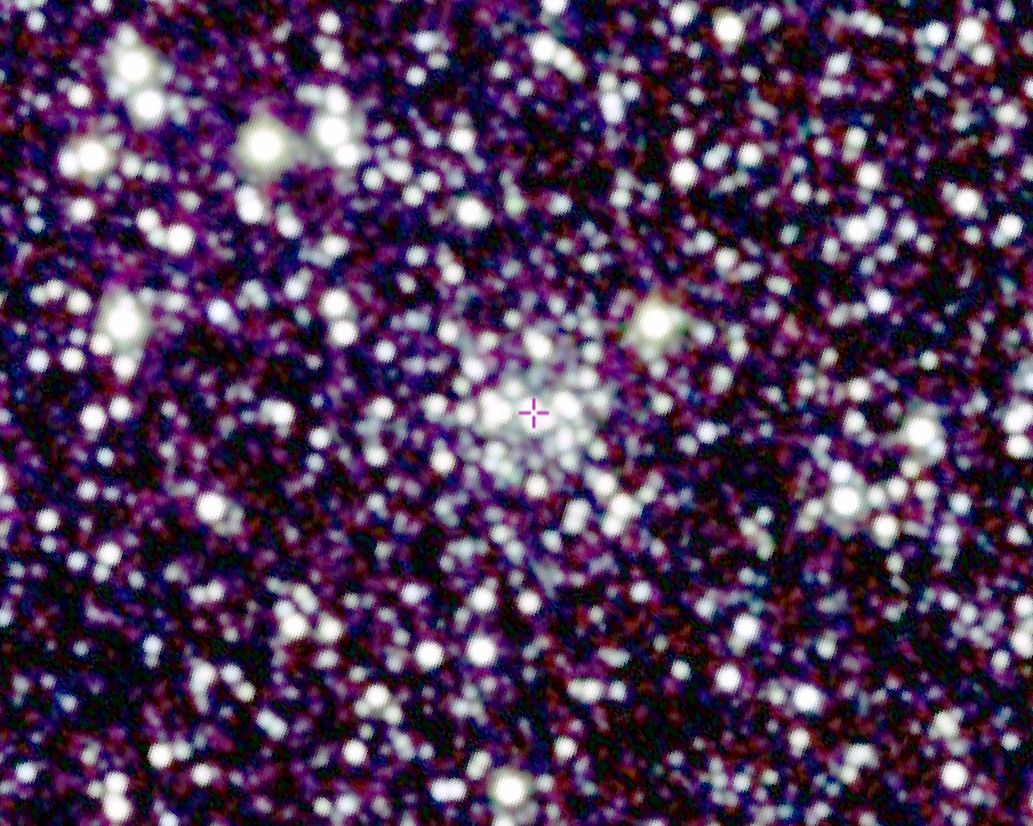}
\includegraphics[height = 4.5cm]{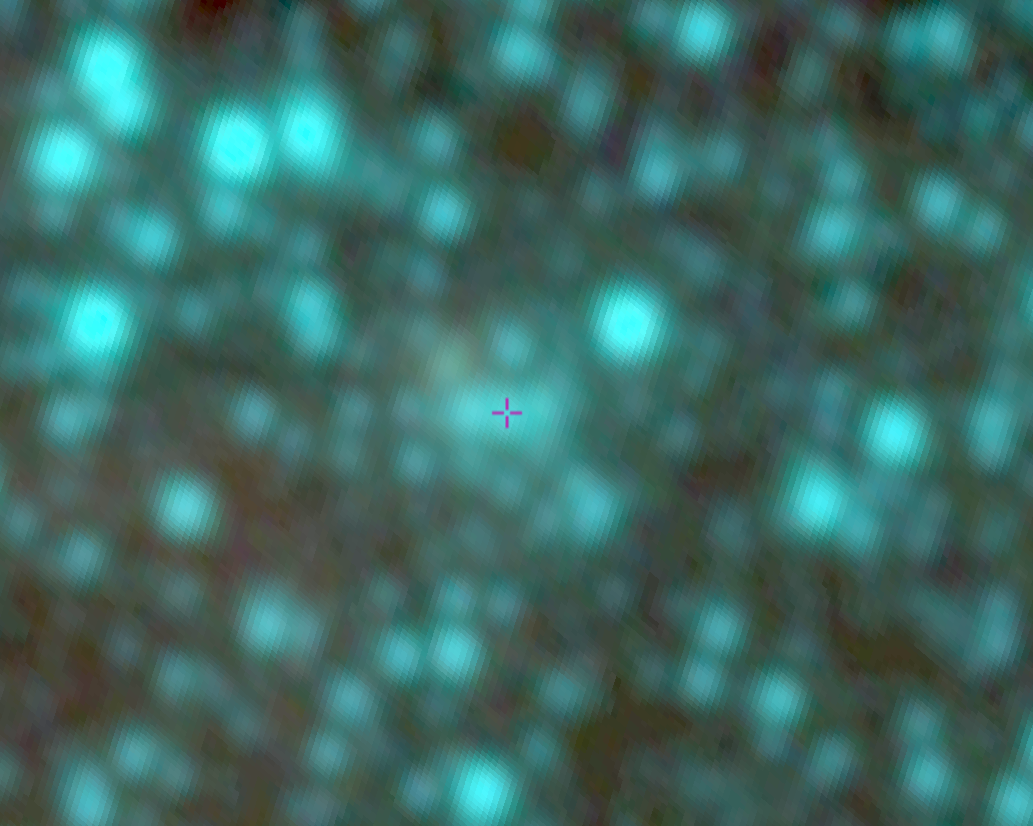}
\includegraphics[height = 4.5cm]{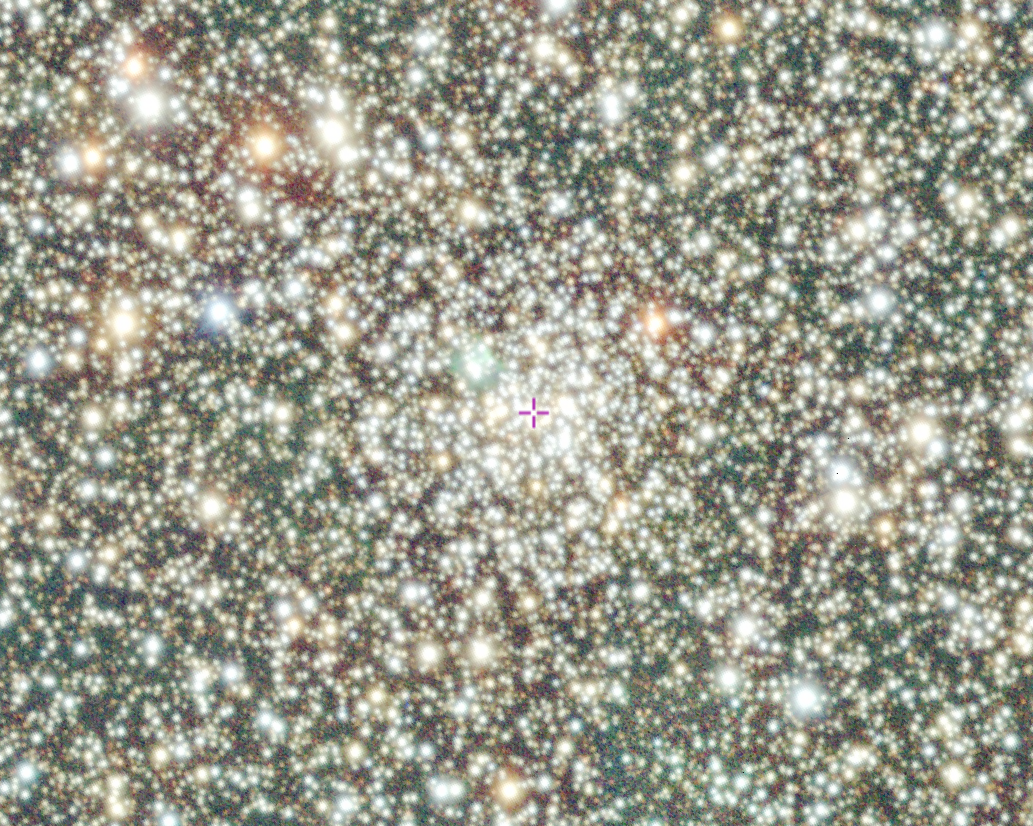}\\
\includegraphics[height = 5cm]{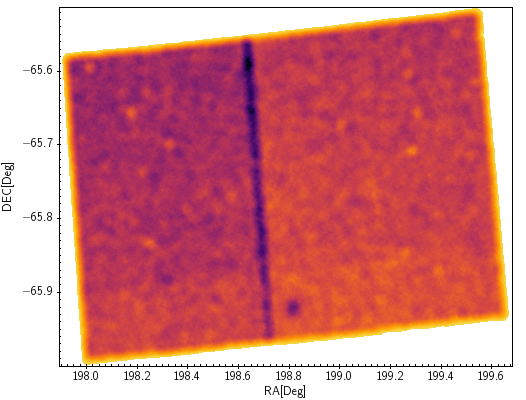}
\includegraphics[height = 5cm]{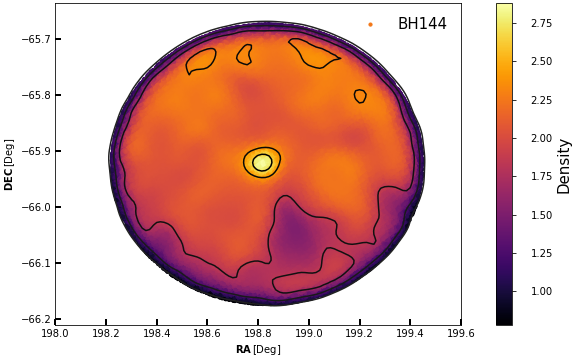}\\

\caption{Four panels are presented: The upper panel shows multi-band images of the open cluster BH118, with the left image from 2MASS, the middle from WISE, and the right from DECaPS. The second panel displays the cluster's density map on the left, highlighting the darkest region as the cluster location, and the KDE distribution on the right. The third and fourth panels present the same data for cluster BH144.}
\label{Figure2A}
\end{center}
\end{figure*}

\section{Methodology}
\label{section3}
To achieve the objectives of this study, we followed methods similar to those outlined in \cite{obasi2021confirmation} and \cite{garro2022new}. We began our investigation by carefully examining images captured at various wavelengths, as shown in Figures \ref{Figure2A} and \ref{Figure2E}. Each figure contains four panels. The first and third panels display cluster images captured at various wavelengths. 
 In all three images in the first and third panels, there is clear evidence of stellar overdensities, which resemble star clusters published in the literature. However, additional checks are required to validate our initial assumption since these overdensities are observed in a densely populated region.

 First, we created a density map using VVVX photometry for the tiles corresponding to each cluster with the TOPCAT program \citep{taylor2005topcat}, revealing a clear excess of stellar density at the cluster location, as shown in the left panels of Figures \ref{Figure2A} and \ref{Figure2E} (second and fourth panels). Starting from the cluster locations in each tile, we applied cone selection and then performed KDE on the resulting data to examine the clusters' overdensities in more detail. This statistical method allows for a more accurate distinction between stellar overdensities and background fluctuations, beyond simple visual inspection. The KDE analysis confirmed a centralized, circular overdensity, as shown in the right panels of Figures \ref{Figure2A} and \ref{Figure2E} (second and fourth panels).
These tests provide compelling evidence that these objects are indeed star clusters. 
\subsection{Decontamination procedure and cluster membership}

\begin{figure}[ht] 
\centering
\includegraphics[height=7cm]{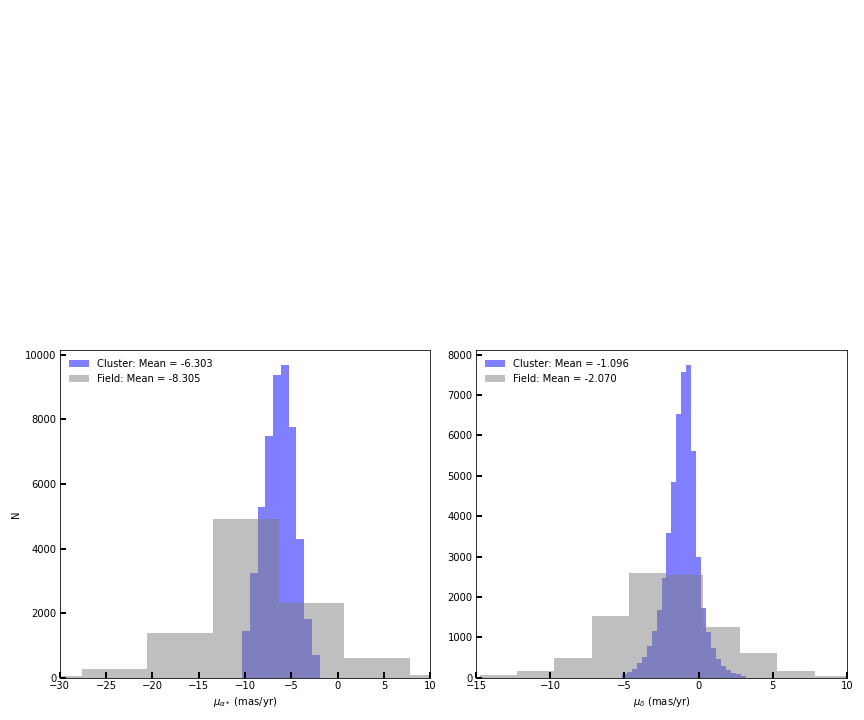}
\caption{The histograms of proper motions (\(\mu_{\alpha*}\) and \(\mu_\delta\)), highlighting the distributions of the cluster (blue) and field (grey) populations along with their respective mean values.}
\label{Figure3AA}
\end{figure}

Studying the inner region of the Milky Way, such as the bulge and disk, is always challenging. Several factors contribute to these difficulties, such as differential reddening and extinction. These factors not only impact the accuracy of photometric distance estimates but also influence derived cluster parameters such as age and metallicity. To mitigate some of these challenges, we applied the following cuts: $\pi$ < 0.5'' (parallax to avoid severe contamination from foreground stars) and renormalized unit weight error (ruwe) < 1.4 (a high-quality astrometric parameter). This ensures that all our photometric data have reliable astrometric solutions which is consistent with our previous studies \citep{garro2022new,garro2024vvvx}.
We then used KDE iso-density contours (Figure \ref{Figure2A} and \ref{Figure2E} second and fourth right panels) to statistically show the significance of the stellar overdensity and eliminate false positives masquerading as genuine stellar overdensities as we previously demonstrated in Figure \ref{Figure1A}. This method is also similar to our analysis in the previous paper \citep{garro2024vvvx}. 
To refine our sample and eliminate contaminants, we computed the membership probability for each cluster. Accurate membership determination is crucial for deriving the clusters' astrophysical parameters. \cite{vasilevskis1958relative} introduced the first mathematical model for membership determination, which was later refined by \cite{sanders1971improved} using the maximum likelihood principle. Since then, improved methods for determining star membership in clusters based on proper motions and different observed precisions have been developed \citep[see][]{stetson1980dwarf,zhao1990improved,zhao1994statistical}. 
We first applied a Gaussian Mixture Model (GMM) with two components to the PM data in RA and Dec to determine cluster membership probabilities, effectively separating cluster stars from field stars. For each group, we calculated the mean and standard deviations of $\mu_{\alpha*}$ and $\mu_{\delta}$.
Furthermore, the separation of the two distributions (D) was quantified by comparing their means $\mu_1$ and $\mu_2$, normalized by their standard deviations $\sigma_1$ and $\sigma_2$. Using D, we can measure how far apart the centres of two distributions are from each other. By doing this, we can determine whether the two distributions overlap or distinct. Our clusters have a minimum D value of 2.5, consistent with the threshold noted by \cite{muratov2010modeling} and \cite{ashman1994detecting}, who suggest that a clean separation between modes requires D>2. Figure \ref{Figure3AA} presents a histogram of the cluster and field distributions for BH144, clearly illustrating the distinct separation of the cluster mean from the field.

We applied a straightforward yet effective method to estimate membership probabilities and identify probable cluster members based on proper motions, closely following the approach of \cite{sanders1971improved}.
 We began by defining the proper motions \( (M_{xi}, M_{yi}) \) as well as their mean values \( (M_{xc}, M_{yc}) \), obtained by using the GMM code, along with their uncertainties \( (E_{M_{xi}}, E_{M_{yi}}) \).   \( \sigma_{c} \) represents the proper motion radius of the cluster. The normalized factor, (factor1), is defined as:
\begin{equation}
\text{factor1} = \sqrt{2 \pi} \cdot \sqrt{\left(\sigma_c^2 + E_{M_{xi}}^2\right) \left(\sigma_c^2 + E_{M_{yi}}^2\right)}
\end{equation}
Next, we calculated the squared distance between each star's proper motion and the cluster mean:

\begin{equation}
\begin{aligned}
    n_{x}^2 &= (M_{xi} - M_{xc})^2 \\
    n_{y}^2 &= (M_{yi} - M_{yc})^2
\end{aligned}
\end{equation}

The variances for \( \text{pmra} \) and \( \text{pmdec} \) are then defined as:

\begin{equation}
\begin{aligned}
    \sigma_{nx}^2 &= \sigma_c^2 + E_{M_{xi}}^2 \\
    \sigma_{ny}^2 &= \sigma_c^2 + E_{M_{yi}}^2
\end{aligned}
\end{equation}

We then computed the exponential component, known as factor2, for the membership probability:

\begin{equation}
\text{factor2} = \exp\left(-\frac{1}{2} \left(\frac{n_{x}^2}{\sigma_{nx}^2} + \frac{n_{y}^2}{\sigma_{ny}^2}\right)\right)
\end{equation}

The unnormalized membership probability \( p_f \) is given by:

\begin{equation}
p_f = \frac{\text{factor2}}{\text{factor1}}
\end{equation}

For clarity, we normalized the membership probabilities to a maximum of 1. This normalization allows for a clear distinction between cluster members and interlopers, defined by:

\begin{equation}
\text{pdf} = \frac{p_f}{\max(p_f)}
\end{equation}

We set our membership threshold \( T \) to classify stars as follows:

\begin{equation}
\text{Membership Status} = 
\begin{cases} 
  \text{Member} & \text{if } p_f \geq T \\ 
  \text{Non-Member} & \text{if } p_f < T 
\end{cases}
\end{equation}

\begin{figure}[ht]
\centering
\includegraphics[height = 8cm]{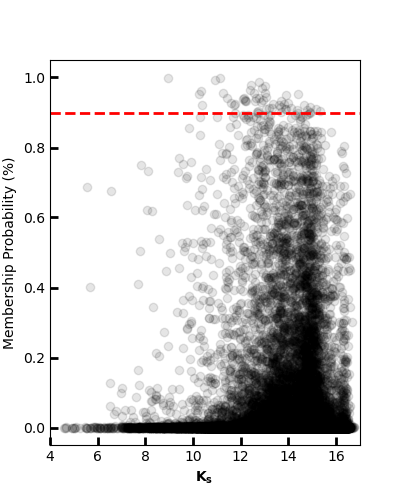}
\caption{Membership probability as a function of K$_{s}$ magnitude is displayed, with a broken horizontal line indicating the 90\% cluster membership threshold for BH144.
}
\label{Figure13A}
\end{figure}

Our selection criteria were strict, with \( T \) typically set to \( \geq 0.9 \) in most cases.
Figure \ref{Figure13A} shows the magnitude vs. membership probability plot for cluster BH144. The red dotted horizontal line indicates a 90\% confidence level that a star belongs to the cluster.

\begin{figure}[ht] 
\centering
\includegraphics[height=5cm]{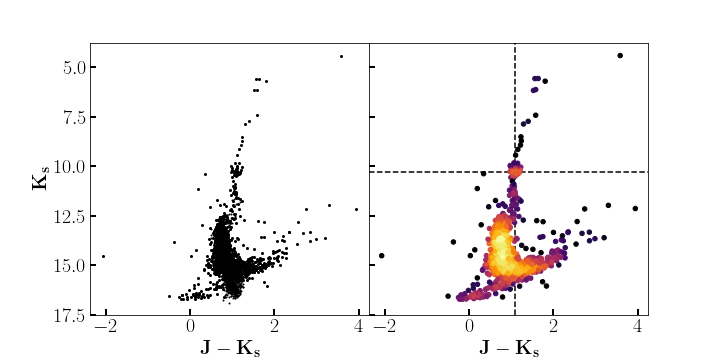}\\
\includegraphics[height=5cm]{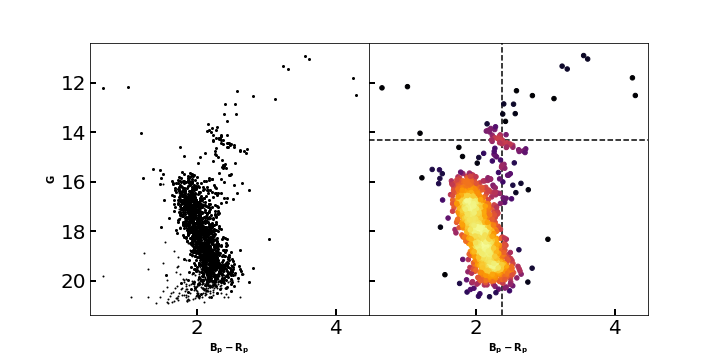}
\caption{In the upper panel, we display the 2MASS NIR CMD (K$_s$ vs. J-K$_s$). On the right, the RC overdensity is visible on the density map, and the horizontal line indicates the density peak. We present the same plot for optical  \textit{Gaia} data (G vs. BP-RP) in the bottom panel for Pismis3.}
\label{Figure4A}
\end{figure}

Subsequently, we constructed the colour-magnitude diagram (CMD) for all the clusters in both the 2MASS NIR and Optical  \textit{Gaia} data. In eight cases, distinct red clump (RC) stars are seen, as observed in Figure \ref{Figure4A} for the cluster Pismis3.
We measured the mean position of the RC overdensity using KDE to compute the max\_density\_index indicated by the dotted horizontal line and their corresponding error. For the clusters listed in Table \ref{table1},  we present in Table \ref{table3} their K$_s$ and J-K$_s$ for the RC.
\begin{table}[h!]
\centering
\caption{Values of $K_{s}$ and $J-K_{s}$ for the Red Clump}

\label{table3}
\begin{tabular}{ccc}
\toprule
Name & $K_{s}$ & $J-K_{s}$ \\
\midrule
BH118 & 13.082 $\pm$ 0.0145 & 0.818 $\pm$ 0.0163 \\
BH\_144 & 13.227 $\pm$ 0.039 & 0.831 $\pm$ 0.082 \\
Schuster-MWSC1756 & 13.255 $\pm$ 0.029 & 2.0 $\pm$ 0.038 \\
Saurer3 & 12.861 $\pm$ 0.025 & 1.031 $\pm$ 0.025 \\
FSR1521 & 11.617 $\pm$ 0.015 & 2.10 $\pm$ 0.045 \\
Saurer2 & 12.398$\pm$0.034 & 1.262 $\pm$ 0.015 \\
Haffner10-FSR1231 & --  & --  \\
Juchert12 & 13.275 $\pm$ 0.01 & 0.919 $\pm$ 0.263 \\
Pismis3 & 10.303 $\pm$ 0.068 & 1.096 $\pm$ 0.079 \\
\bottomrule
\end{tabular}
\end{table}

\section{Estimation of the main physical parameters}
\label{section4}
We used the final catalogue obtained from our decontamination process for the rest of the analysis. We derived distance, reddening, and extinction. We built CMDs of the clusters and derived astrophysical parameters such as age and metallicity. The use of multiband photometry in our analysis, comprising VVVX+2MASS and Optical  \textit{Gaia}, helped us obtain more robust results. The structural parameters such as core radii, tidal radii and concentration parameters are presented in the next subsections.
\subsection{Reddening, extinction, distance, vertical component and tangential velocity}
\label{section4.1}
We computed the reddening and extinction associated with each cluster we studied. To do this, we followed the method outlined by \cite{ruiz2018empirical}, as we have done in our previous studies \citep{obasi2021confirmation,garro2022new}. We used two methods to verify the position of the red clump stars in each cluster. First, we built the CMD (J vs. J-K$_s$) and used KDE to estimate the positions of K$_s$ and J-K$_s$ as demonstrated in Figure \ref{Figure4A} (upper panel). We then verified the consistency of this method by constructing the generic luminosity function in the K$_s$ band for each cluster, finding that the values obtained were in agreement.

 Three methods were used to calibrate the extinction and reddening in our sample. First, we used \cite{schlafly2011measuring} and \cite{cardelli1989relationship} to determine the extinction and reddening values. Second, we applied the extinction law from \cite{nataf2015interstellar}. Third, we calibrated our extinction and reddening values using the absolute magnitude of RC stars in the K$_s$ band (\(M_{K_s} = -1.605 \pm 0.009\)) and their intrinsic colour (\((J - K_s)_0 = 0.66 \pm 0.02\)) from \citep{ruiz2018empirical}. 
Based on these methods, we reported only the values from \cite{nataf2015interstellar} and \cite{ruiz2018empirical}, as they showed excellent agreement in their derived distances. We adopted the \cite{ruiz2018empirical} extinction law for the corrections. In Section \ref{section6}, we will briefly discuss the reasons for the differences between these results and the exclusion of   \cite{schlafly2011measuring}.

For the young cluster Haffner 10, whose CMD lacks a well-populated red RC sequence due to its age ($\sim$ 20 Myrs), we applied the extinction law derived by \cite{nishiyama2009interstellar} (A$_{K_s}$=0.67$\cdot$ E(J - K$_{s}$)
 to compute both extinction and reddening.  The derived extinction for all the clusters varied widely, ranging from A$_{ks}$= 0.07$\pm$0.03  to 0.50$\pm$0.04 , and the reddening ranged from E(J-Ks)$_0$= 0.16$\pm$0.03  to 0.60$\pm$0.03 . Using these values, we estimated the heliocentric distances of the clusters by applying the distance modulus equation. For Haffner 10, we first computed its distance using \textit{Gaia} parallaxes.  Both methods yielded similar distance results for each cluster; therefore, we adopted and reported the average values in table \ref{table2}, which range from D =  2.19$\pm$0.06 kpc to D =  8.94$\pm$0.06 kpc.

We adopted R$_\odot$ = 8.2 kpc \citep{abuter2019geometric} to compute the Galactocentric distance, which ranged from R$_G$=  7.82 kpc to  15.08 kpc. Using the relation Z = D$\times$sin(b), assuming Z$_\odot$ = 0.007 kpc \citep{siegert2019vertical}, we estimated the distance to the Galactic plane for each cluster, which ranged from Z= -0.09 kpc to 0.34 kpc, both below and above the Galactic plane. Finally, we computed the tangential velocities of each cluster, which ranged from V$_T$= 30.59 km/s to  245.42 km/s. This kinematic parameter allows us to deduce the Galactic component each of the clusters belongs to, which we will discuss further in the subsequent section.
\subsection{Ages and Metallicities}
\label{section4.2}
Following a similar procedure to our previous work \citep{obasi2021confirmation} and taking into account our derived reddening values, extinctions, and distances for each cluster, we fitted PARSEC isochrones \citep{bressan2012parsec}. We visually inspected the fit until a match was obtained. 
Age and [Fe/H] were then obtained from this procedure.

We fixed the reddening and distance values estimated from the RC stars and fit the isochrones for different ages and metallicities. To derive robust fits, we varied the age and metallicity until an optimal match was found. This method reliably constrained the age and [Fe/H] for each cluster, as demonstrated in Section \ref{section6}, where we discuss our results and compare them to values in the literature. The best-fitted isochrones in the VVVX and  \textit{Gaia} CMDs are shown in Figures \ref{Figure5A} and \ref{Figure5B}-\ref{Figure5E}. The best-fitted parameters are listed in Table \ref{table2}. The uncertainties in metallicities and distances in our study are estimated to be approximately 0.5 dex in ages and 0.2 dex in metallicities, consistent with findings from similar studies on open clusters \citep[e.g.;][]{garro2024vvvx,dias2024three}.

\subsection{Structural parameters}
\label{section4.3}
We measured the physical size of the clusters following a similar approach used by \cite{obasi2021confirmation}. First, we computed the Radial Density Profile (RDP). The initial centre coordinates obtained from the literature for each cluster were recomputed. The new centre was determined by calculating the median values of $\mu_{\alpha}$ and $\mu_{\delta}$ within a radius of 0.1' from the literature values. This newly derived centre was then used to build the RDPs.

For each cluster, we divided our sample into circular annuli with increasing radii moving outward from the centre. The number density per bin was computed as the number of stars (N) in the bin divided by the area (A). The RDP profile for each cluster is plotted as a function of the mean distance of the circular annulus from the cluster centre against the number density in the corresponding annuli. The error bars were computed as \( e = \sqrt{N/A} \), following the method used in our previous work \citep{garro2022new}. We subtracted the background level, which varies from 0.3 to 0.8 stars/arcmin\(^2\), depending on the cluster.

To derive the physical parameters of the clusters, we used the model by \cite{king1962structure} to fit the cluster density profile. 
From the best-fit parameters, we obtained the core radius (\(r_c\)), which ranges from 0.71' to 5.21'; the tidal radius (\(r_t\)), varying from 3.6' to 12'; and the concentration parameter (c), which ranges from 0.36' to 0.83'. See Figure \ref{Figure7} for details.

\section{Results}
\label{section5}

 We employed a multiwavelength approach using VVVX+ \textit{Gaia} DR3 and 2MASS+ \textit{Gaia} DR3 to improve measurements, resolve inconsistencies from previous literature, and provide a detailed characterization of Jurchert12 for the first time. This study encompasses nine open clusters.

 We began by deriving the extinction and reddening values for each cluster using the relation provided by \cite{nataf2015interstellar}, where A$_V$/A$_{Ks}$ = 13.44 and E(J-K$s$)/A$_{Ks}$ = 0.43. This relation is optimized for the Milky Way's inner region, where extinction levels are particularly high. The results obtained from this method were compared with those from \cite{ruiz2018empirical} to ensure consistency. The A$_V$ values for each cluster were retrieved from the NASA/IPAC database\footnote{\url{https://irsa.ipac.caltech.edu/cgi-bin/bgTools/nph-bgExec}}, which uses the relation from \cite{schlafly2011measuring} to compute A$_V$.

For BH118, we found  A$_V$ = 1.66 $\pm$ 0.10   and  E(B-V) = 0.53 $\pm$ 0.05  . For BH144, we found  A$_V$ = 2.52 $\pm$ 0.10  and  E(B-V) = 0.81 $\pm$ 0.06  . The results for the other clusters are presented in Table \ref{tables4} and \ref{table2}.

 Next, we applied the extinction and reddening relation by \cite{nataf2015interstellar} to determine each cluster's NIR extinction and reddening. For BH118, we obtained A$_{K_s}$ = 0.12 $\pm$ 0.01  and E(J-K$_s$) = 0.05 $\pm$ 0.0. For BH144, we found A$_{K_s}$ = 0.19 $\pm$ 0.01  and E(J-K$_s$) = 0.08 $\pm$ 0.0. The values for the other clusters are presented in table \ref{tables4}.

Using these parameters and the absolute M$_K$ derived by \cite{alves2002k} M$_K$=-1.61$\pm$0.03, we estimated the distance modulus (\(m-M\))$_0$ for each cluster: 14.57 $\pm$ 0.03  for BH118, 14.65 $\pm$ 0.05  for BH144. We have reported the results for other clusters in table \ref{tables4}.

These values translate to heliocentric distances of  8.20 $\pm$ 0.13 kpc for BH118, 8.53 $\pm$ 0.20 kpc for BH144. The results for other clusters are presented in table \ref{tables4}.

We then validated these values using the \cite{ruiz2018empirical} relation. After measuring the RC positions in each cluster. To compute for the extinction we adopted the relation given in \cite{alonso2017extinction} (A$_{K_s}$/(E(J-K$_s$)=0.428$\pm$0.005) we found the following NIR extinction and excess colour values: for BH118, \(A_{K_s} = 0.07 \pm 0.03\)  and \(E(J-K_s) = 0.16 \pm 0.03\); for BH144, \(A_{K_s} = 0.07 \pm 0.08\)  and \(E(J-K_s) = 0.17 \pm 0.08\). We have presented the results for the other clusters in table \ref{tables4}.
 These values were used to estimate the distance modulus and distances to the clusters, resulting in the following: for BH118, \(14.61 \pm 0.03\)  and \(8.36 \pm 0.03\) kpc; for BH144, \(14.76 \pm 0.11\)  and \(8.95 \pm 0.11\) kpc. The other cluster's results are listed in table \ref{tables4}.

\begin{figure}[ht] 
\centering
\includegraphics[height=5cm]{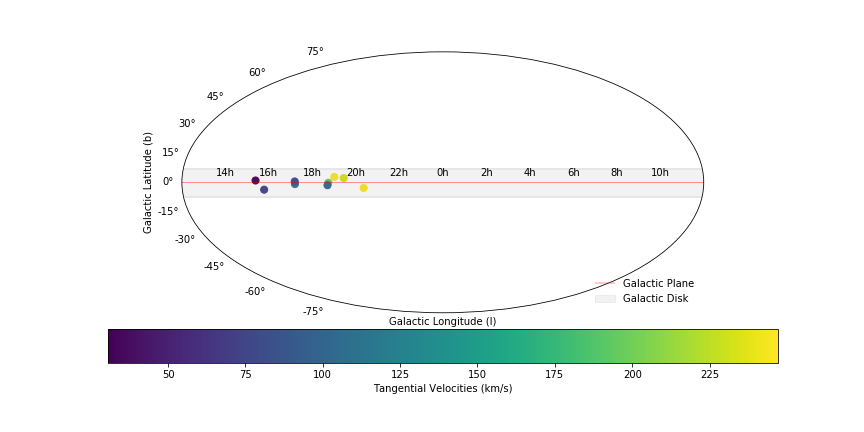}
\caption{A sky distribution plot of all nine clusters, showing their positions in Galactic coordinates (longitude $l$  and latitude  $b$, with colour representing their tangential velocities.}
\label{FigureZ}
\end{figure}

 Although extinction and reddening were calculated differently using the methods from \cite{nataf2015interstellar} and \cite{ruiz2018empirical}, both approaches yielded distance results in excellent agreement
. We averaged the distances from these methods (see Table \ref{table2}) and compared them with those obtained using \textit{Gaia} data. For \textit{Gaia}, we used the \cite{babusiaux2018gaia} relation to compute extinction A$_G$ and colour excess Bp-Rp. Since higher extinction affects distance calculations, we only reported distances for clusters with A$_G$ $<$ 4 . The \textit{Gaia} distances and the A$_G$ value from the \textit{Gaia} data are presented in Table \ref{table2}, and the \textit{Gaia} distances are within 2$\sigma$ of the other distances.
 Furthermore, we used the new average distances to calculate the Galactocentric distance R$_G$ for each cluster, adopting R$_{\odot}$ = 8.2 kpc \citep{abuter2019geometric}. The R$_G$ values obtained ranged from  7.82 kpc to  15.08 kpc (see Table \ref{table2} for details). Furthermore, we computed the vertical distance Z, which ranged from -0.09 to 0.34 kpc for all clusters.  Figure \ref{FigureZ} shows the sky distribution of the clusters according to vertical distance (see Table \ref{table2} for the complete Z values). In Table \ref{table2}, the computed proper motion and tangential velocities for each cluster are also shown.

Finally, using isochrone fitting, we estimated the ages and [Fe/H] values for all clusters. The ages range from 20 Myrs to 5 Gyrs, while the [Fe/H] values range from -0.5 to 0.5 dex. The results of the isochrone fitting are presented in Table \ref{table2}. Additionally, we derived structural parameters for the clusters, including core radius (\(r_c\)), tidal radius (\(r_t\)), and concentration index (\(c\)). The structural parameter values, also shown in Table \ref{table2}, confirm that all the clusters are genuine open clusters and none have undergone core collapse.

\begin{table*}[h!]
\centering
\caption{Extinction, reddening, distance modulus, and distances for the clusters.}
\begin{tabular}{lcccccc}
\midrule
\textbf{ Nataf method} & & & & & & \\ 
\toprule
Name & A$_V$  & E(B-V)  & A$_{K_s}$  & E(J-K$_s$)  & (m-M)$_0$  & D (kpc) \\ 
\midrule
BH118 & 1.66$\pm$0.10 & 0.53$\pm$0.05 & 0.12$\pm$0.01 & 0.05$\pm$0.00 & 14.57$\pm$0.02 & 8.18$\pm$0.07 \\
BH144 & 2.52$\pm$0.10 & 0.81$\pm$0.06 & 0.19$\pm$0.01 & 0.08$\pm$0.00 & 14.64$\pm$0.04 & 8.51$\pm$0.16 \\
Schuster-MWSC1756 & 10.76$\pm$0.10 & 3.47$\pm$0.20 & 0.80$\pm$0.01 & 0.34$\pm$0.01 & 14.06$\pm$0.03 & 6.49$\pm$0.10 \\
Saurer3 & 2.56$\pm$0.10 & 0.83$\pm$0.06 & 0.19$\pm$0.01 & 0.08$\pm$0.00 & 14.28$\pm$0.03 & 7.16$\pm$0.09 \\
FSR1521 & 10.32$\pm$0.10 & 3.33$\pm$0.20 & 0.77$\pm$0.01 & 0.33$\pm$0.01 & 12.45$\pm$0.02 & 3.09$\pm$0.03 \\
Saurer2 & 5.85$\pm$0.10 & 1.89$\pm$0.11 & 0.43$\pm$0.01 & 0.19$\pm$0.04 & 13.56$\pm$0.04 & 5.14$\pm$0.09 \\
Haffner10-FSR1231 & 2.05$\pm$0.10 & 0.66$\pm$0.05 & 0.15$\pm$0.01 & 0.07$\pm$0.00 & 12.66$\pm$0.04 & 3.62$\pm$0.07 \\
Jurchert12 & 2.60$\pm$0.10 & 0.84$\pm$0.06 & 0.19$\pm$0.01 & 0.08$\pm$0.00 & 14.69$\pm$0.02 & 8.84$\pm$0.07 \\
Pismis3 & 4.37$\pm$0.10 & 1.41$\pm$0.09 & 0.33$\pm$0.09 & 0.14$\pm$0.04 & 11.72$\pm$0.07 & 2.20$\pm$0.07 \\ 
\midrule
\textbf{Ruiz-Dern method} & & & & & & \\ \midrule
BH118 & & & 0.07$\pm$0.03 & 0.16$\pm$0.03 & 14.61$\pm$0.03 & 8.36$\pm$0.03 \\
BH144 & & & 0.07$\pm$0.08 & 0.17$\pm$0.08 & 14.76$\pm$0.11 & 8.95$\pm$0.11 \\
Schuster-MWSC1756 & & & 0.57$\pm$0.04 & 1.34$\pm$0.04 & 14.29$\pm$0.05 & 7.21$\pm$0.05 \\
Saurer3 & & & 0.16$\pm$0.03 & 0.37$\pm$0.03 & 14.30$\pm$0.04 & 7.25$\pm$0.04 \\
FSR1521 & & & 0.62$\pm$0.05 & 1.44$\pm$0.05 & 12.59$\pm$0.10 & 3.31$\pm$0.10 \\
Saurer2 & & & 0.26$\pm$0.03 & 0.60$\pm$0.03 & 13.74$\pm$0.05 & 5.59$\pm$0.05 \\
Haffner10-FSR1231 &--&-- & -- &-- &-- &--\\
Jurchert12 & & & 0.11$\pm$0.07 & 0.26$\pm$0.07 & 14.77$\pm$0.07 & 8.97$\pm$0.07 \\
Pismis3 & & & 0.19$\pm$0.08 & 0.44$\pm$0.08 & 11.71$\pm$0.10 & 2.19$\pm$0.10 \\ 
\bottomrule
\end{tabular}
\label{tables4}

\textbf{Note:} The Nataf method derives from the relation presented in \cite{nataf2015interstellar}. However, we obtained the A$_V$ values from the NASA/IPAC Infrared Science Archive (\url{https://irsa.ipac.caltech.edu/applications/DUST/}), which uses the extinction maps from \cite{schlafly2011measuring} and \cite{schlegel1998maps} to compute A$_V$ for a given source. For our analysis, we adopted the values provided by SF11. It is important to note that the errors in A$_{Ks}$ and E(J-K$_s$) are not zero. When rounded to three decimal places, the values are A$_{K_s}$ = \textit{[ 0.008,0.008,0.013,0.008,0.013,0.010,0.010,0.086]} and E(J-K$_s$) = \textit{[ 0.003,0.003,0.006,0.003,0.006,0.041,0.003,0.003,0.037].}

\end{table*}

\section{Discussion }
\label{section6}
A discussion of our results is presented in this section, along with a comparison of our obtained values with those found in the literature where available. In order to improve previous measurements and study the cluster in detail, we used a multiwavelength approach using VVVX+\textit{Gaia} DR3 and 2MASS+\textit{Gaia} DR3. This discussion begins by comparing each object with literature values, followed by an observation of the general trends observed across the sample.

 BH118: For the cluster BH118, we obtained   A$_V$ = 1.66 $\pm$ 0.10   and   E(B-V) = 0.53 $\pm$ 0.05. \cite{cantat2020painting}, however, reported a lower extinction value of A$_V$ = 1.07. Using both Nataf methods and the Ruiz-Dern method for NIR, we obtained the following values:   A$_{Ks}$ and E(J-K$_s$)  as 0.12 $\pm$ 0.01, 0.05 $\pm$ 0.003  and 0.07 $\pm$ 0.03,  0.16 $\pm$ 0.03, respectively. With these parameters, we computed a distance modulus of   14.57 $\pm$ 0.03  and  14.61 $\pm$ 0.03, equivalent to distances of  8.20 $\pm$ 0.13 kpc and  8.36 $\pm$ 0.03 kpc for both methods.  Both results are in good agreement with those obtained by \cite{cantat2020painting}, where (m-M)$_0$ = 14.72 and D = 8.8 kpc. The Galactocentric distance  R$_G$ of 9.20 kpc we computed for BH118 is consistent with the 9.6 kpc value obtained by \cite{cantat2020painting}. Additionally, the vertical distance of 0.34 kpc we obtained aligns with the 0.36 kpc reported by \cite{cantat2020painting}.

\begin{figure*}[ht] % Changed to [ht] for more flexibility in placement
\centering
\includegraphics[height=10cm]{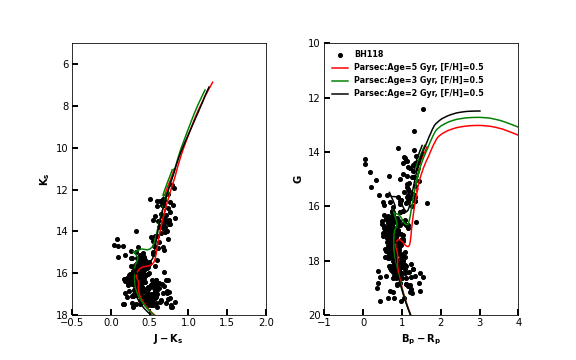}
\caption{Colour-magnitude diagrams for BH118 cluster studied in both NIR and optical  \textit{Gaia} bands. These CMDs are fitted with Parsec isochrones to infer the age and [Fe/H] for each cluster. }
\label{Figure5A}
\end{figure*}

Our computed proper motion values  $\mu$ $_\alpha$ = -5.821 $\pm$ 0.209 mas yr$^{-1}$ and $\mu$ $_\delta$ = 1.609 $\pm$ 0.280 mas yr$^{-1}$ are in good agreement with those computed by \cite{cantat2020painting}, which are  $\mu$ $_\alpha$ = -4.406  mas yr$^{-1}$ and $\mu$ $_\delta$ = 1.450 mas yr$^{-1}$. Furthermore, we derived the tangential velocities using these proper motions and distances as  V$_{T\alpha*}$ $\approx$ -225.42 km/s and  V$_{T\delta}$ $\approx$ 62.31  km/s. These positional and kinematic parameters suggest that BH118 is a disk cluster, as we will discuss later in this section. Finally, we derived an age of  3 $\pm$ 0.5  Gyr and [Fe/H] = -0.5 dex . However, \cite{cantat2020painting} estimated the age of BH118 to be $\approx$ 2 Gyr. We over-plotted the isochrone family of 5, 3, and 2 Gyr to the CMD, as shown in Figure \ref{Figure5A}. The 3 Gyr isochrone provides a much better fit to the cluster. The core radius  (r$_c$ ), tidal radius ( r$_t$ ), and concentration index (c) of 0.86 $\pm$ 0.08', 5.76 $\pm$ 0.92' (see Figure \ref{Figure7}), and 0.83, respectively, all indicate a typical open cluster found in the disk \citep[see][]{garro2022new,dias2024three}.

{\bf BH144:} For BH144, our derived  A$_V$ value of  2.52 $\pm$ 0.10   is slightly higher than the value provided by \cite{cantat2020painting}  A$_V$ = 1.68  . However, our colour excess  E(B-V) = 0.81 $\pm$ 0.06   is in good agreement with the value of 0.63  obtained by \cite{kharchenko2013global}. Using the two methods previously described, we obtained  A$_{Ks}$  and  E(J-K$_s$) values of   0.19 $\pm$ 0.01 ,  0.08 $\pm$ 0.0   and  0.07 $\pm$ 0.08  ,  0.17 $\pm$ 0.08  , respectively. The distance we computed using the Ruiz-Dern method, 8.95 kpc, agrees with the \cite{cantat2020painting} estimation of 9.6 kpc. Furthermore, our computed Galactocentric distance  R$_G$ =  7.82 kpc aligns with the value calculated by \cite{cantat2020painting} of 8.3 kpc. Our calculated vertical distance of  -0.49 kpc also agrees with the value of -0.53 kpc reported by \cite{cantat2020painting}. Our proper motion values are in excellent agreement with those of \cite{cantat2020painting} (see Table \ref{table2} for comparison). However, the values computed by \cite{kharchenko2013global}  $\mu$ $_\alpha$ = -13.19  mas yr$^{-1}$ and  $\mu$ $_\delta$ = 7.74  mas yr$^{-1}$ largely deviate from our calculations. Similarly, the values provided by \cite{dias2014proper}  $\mu$ $_\alpha$ = -4.73 mas yr$^{-1}$ and  $\mu$ $_\delta$ = -3.36 mas yr$^{-1}$ also deviate slightly from our values. Our derived tangential velocities for BH144 (see Table \ref{table2}) suggest that this cluster is located in the disk. Finally, isochrones of 6, 4, and 2 Gyr were utilized, with the best fit at  2 $\pm$ 0.2  Gyr and [Fe/H] = 0.5 dex. \cite{kharchenko2013global} estimated an age of 1 Gyr with [Fe/H] = -0.5 dex, while \cite{cantat2020painting} derived an age of 1.5 Gyr, consistent with our findings (see Figure \ref{Figure5B}). The structural parameters  r$_c$  and  r$_t$  of 1.9 pc and 11.88 pc slightly differ from \cite{kharchenko2013global} values of 0.83 pc and 12.88 pc.

{\bf Schuster-MWSC1756:} We calculated A$_V$ = 10.76 $\pm$ 0.10  and E(B-V) = 3.47 $\pm$ 0.20   for the cluster Schuster-MWSC1756. In contrast, \cite{kharchenko2013global} calculated a lower colour excess of  E(B-V) = 1.08  . In Table \ref{tables4}, we report our NIR  A$_{K_s}$, E(J-K$_s$) values, the distance modulus, and cluster distance,  while in Table \ref{table2} we report the Galactocentric distance based on these parameters.

Our computed proper motion values are in good agreement with those computed by \cite{cantat2020painting} (see Table \ref{table2}), but they differ significantly from the values obtained by \cite{kharchenko2013global} $\mu$ $_\alpha$ = -9.41  mas yr$^{-1}$ and $\mu$ $_\delta$ = 3.06  mas yr$^{-1}$. 
The tangential velocities suggest that this cluster lies in the disk. Of all the clusters we examined, Schuster-MWSC175 is the smallest size, with core and tidal radii of \( r_c = 1.4 \) pc and \( r_t = 7.04 \) pc, respectively (See Figure\ref{Figure7}). Given the high extinction in this cluster (\( A_V > 10 \)), effective decontamination was challenging, particularly in the \textit{ Gaia} optical CMD. We compared isochrone fits at 400, 700, and 900 Myr with [Fe/H] of -0.5,0.0 and 0.0 dex (see Figure \ref{Figure5C}); however, the sparsity of the CMD data makes it difficult to precisely determine the cluster’s age, which tends to be at 700 Myrs. Nevertheless, it appears significantly older than the 6.3 Myr estimated by \cite{kharchenko2013global}. For clusters with A$_V$ $\geq$ 5, we did not derive distances using Gaia data, and therefore we did not fit any isochrones on them.

{\bf Saurer 3:} The E(B-V) = 0.83 $\pm$ 0.06   value we obtained for Saurer 3 is consistent with the 0.78  derived by \cite{kharchenko2013global}. However, our computed proper motion values,  $\mu$ $_\alpha$ = -6.62 $\pm$ 0.20 mas yr$^{-1}$ and  $\mu$ $_\delta$ = 3.24 $\pm$ 0.91  mas yr$^{-1}$, differ from those computed by \cite{dias2014proper}, who estimated $\mu$ $_\alpha$ = -6.73  mas yr$^{-1}$ and $\mu$ $_\delta$ = -1.62 mas yr$^{-1}$. These values are also significantly different from those obtained by \cite{kharchenko2013global}, which are  $\mu$ $_\alpha$ = -8.34  mas yr$^{-1}$ and  $\mu$ $_\delta$ = 4.40  mas yr$^{-1}$.
Furthermore, our derived age of 2 $\pm$ 0.2 Gyr and [Fe/H] = 0.5 dex (See Figure \ref{Figure5C} bottom panel) aligns closely with the $\approx$ 2 Gyr estimate reported by \cite{kharchenko2013global}.
We also derived structural parameters: a  r$_c$  of  0.79 $\pm$ 0.12' (corresponding to ( r$_c$ = 1.6 ) pc), a  r$_t$ of  4.65 $\pm$ 0.21'  (corresponding to ( r$_t$ = 9.6  pc), and a concentration index (c) of 0.77. These values are significantly different from those provided by \citep{kharchenko2013global}, which are  r$_c$ = 9.18  pc and  r$_t$ = 20.46  pc, respectively.

{\bf FSR 1521:} \cite{cantat2020painting} obtained an A$_V$ of 2.72, which is about eight magnitudes smaller than the A$_V$=10.32 $\pm$ 0.10 that we obtained for the cluster FSR 1521 using the extinction relation from \cite{schlafly2011measuring}. Additionally, our E(B-V) = 3.33 $\pm$ 0.20  differs significantly from the 0.37  measured by \cite{kharchenko2013global}. These differences between our values and those in the literature might stem from our approach to computing optical extinction, where we use values provided by \cite{schlafly2011measuring}.

Our computed (m - M)$_0$ =  12.46 $\pm$ 0.03  and distance of  3.10 $\pm$ 0.05 kpc agree reasonably well with (m - M)$_0$ = 13.47  and D = 4.9 kpc reported by \cite{cantat2020painting}. We computed a Galactocentric distance R$_G$ =  8.23 kpc, which is consistent with the 8.8 kpc calculated by \cite{cantat2020painting}. Similarly, our vertical-horizontal distance is consistent with the value provided by \cite{cantat2020painting} (see Table \ref{table2} for comparison). Our proper motion value is also in agreement with those obtained by \cite{cantat2020painting} (see Table \ref{table2} for comparison). However, the value -8.92  mas yr$^{-1}$, 6.10  mas yr$^{-1}$ provided by \cite{kharchenko2013global} significantly differs from our computed value.

Using isochrone fitting, we estimated the cluster's age to be 1 $\pm$ 0.2 Gyr and [Fe/H] = 0.0 dex, as shown in the upper panel of Figure \ref{Figure5D}. This age estimate differs significantly from previous studies: \cite{cantat2020painting} reported an age of 5.2 Gyr, which appears to be an overestimate, while \cite{kharchenko2013global} calculated an age of 5 Myr, likely underestimating the cluster's age. We also derived structural parameters: a  r$_c$ of 2.19 $\pm$ 0.27' (corresponding to r$_c$ = 2.5 pc), a r$_t$ of 7.69 $\pm$ 0.92' (corresponding to r$_t$ = 8.9 pc), and a concentration index c of 0.55. However, our measured values significantly differ from those obtained by \cite{kharchenko2013global}, who reported r$_c$ = 0.84 pc and r$_t$ = 6.69 pc, respectively.

{\bf Saurer 2:}
We obtained A$_V$ = 5.85 $\pm$ 0.10  for the cluster Saurer 2. This value contrasts with the A$_V$ = 3.72  derived by \cite{cantat2020painting}, while our colour excess E(B-V) = 1.89 $\pm$ 0.11 is consistent with the 1.25  derived by \cite{kharchenko2013global}. Our computed distance of  5.15 $\pm$ 0.11 kpc is in good agreement with the value of D = 6.4 kpc obtained by \cite{cantat2020painting}. Furthermore, our computed Galactocentric distance R$_G$ =  10.77 kpc also aligns well with the R$_G$ = 12.1 kpc obtained by \cite{cantat2020painting}.

 Our measured Saurer 2 vertical-horizontal distance of  -0.10 kpc is consistent with the -0.13 kpc value computed by \cite{cantat2020painting}. In the same way, our computed proper motion agrees well with that of \cite{cantat2020painting} (see Table \ref{table2} for comparison), but it differs significantly from the values reported by \cite{kharchenko2013global} (see also Table \ref{table2}).
Using isochrone fitting, we estimated the cluster's age to be 700 $\pm$ 0.2 Myr with [Fe/H] = 0.5 dex  (See Figure \ref{Figure5C} upper left panel). This estimate contrasts with the ages of 1.6 Gyr and 1.8 Gyr reported by \cite{cantat2020painting} and \cite{kharchenko2013global}, respectively, both of which likely overestimate the cluster’s age. The structural parameters derived for Saurer 2 r$_c$ = 1.4 pc, r$_t$ = 7.8 pc (See Figure \ref{Figure7}) are in good agreement with those measured by \cite{kharchenko2013global} r$_c$ = 1.4 pc and r$_t$ = 15.56 pc.

{\bf Haffner10-FSR1231:}
There are a few parameters available for comparison with Haffner 10. However, the distance of 3.4 kpc computed by \cite{cantat2020painting} is in good agreement with the  3.63 $\pm$ 0.09 kpc we obtained. The proper motion we computed is also in good agreement with those provided by \cite{cantat2020painting}  and \cite{dias2014proper} (see Table \ref{table2} for comparison).
 We estimated an age of 20 $\pm$ 0.2 Myr and [Fe/H] = 0.5 dex for the cluster Haffner 10, which is notably different from the 3.8 Gyr age reported by \cite{cantat2020painting}. Our result indicates that Haffner 10 is a much younger open cluster than previously thought. (see Figure \ref{Figure5D} bottom panel). The other parameters derived for this cluster are listed in Tables \ref{tables4} and \ref{table2}.

{\bf Juchert 12:}
Juchert 12 is an open cluster catalogued by \cite{kronberger2006new}. Its J2000 coordinates are $\mu$ $_\alpha$ = 07:20:56.712 and $\mu$ $_\delta$ = -22:52:00.12, with Galactic coordinates $l$ = 236.5614 and $b$ = -4.1232. To the best of our knowledge, this cluster has not been previously characterized. The only parameter available in the literature is the proper motion value estimated by \cite{dias2014proper}, who reported $\mu$ $_\alpha$ = -3.46 mas yr$^{-1}$ and $\mu$ $_\delta$ = 0.52 mas yr$^{-1}$. These values contrast with our findings, which are $\mu$ $_\alpha$ = -0.75 $\pm$ 0.026 mas yr$^{-1}$ and $\mu$ $_\delta$ = 1.62 $\pm$ 0.051 mas yr$^{-1}$.  
In this analysis, we determined the physical parameters of this cluster for the first time. Through isochrone fitting, we estimated an age of 2 Gyr and an [Fe/H]=0.5 dex (See Figure \ref{Figure5E} upper panel). Additional parameters for this cluster are presented in Tables \ref{tables4} and \ref{table2}.

{\bf Pismis 3:}
The A$_V$ = 2.35  obtained by \cite{cantat2020painting} is lower than the 4.37 $\pm$ 0.10  we derived from our analysis. We computed a distance of  2.19 $\pm$ 0.06 kpc for the cluster, which agrees with the 2.35 kpc estimated by \cite{cantat2020painting}. Additionally, our computed proper motion values $\mu$ $_\alpha$ = -4.76 $\pm$ 0.38 mas yr$^{-1}$ and $\mu$ $_\delta$ = 6.66 $\pm$ 0.01 mas yr$^{-1}$ agree with those provided by \cite{cantat2020painting} (see Table \ref{table2} for comparison). However, these values slightly differ from those computed by \cite{dias2014proper} $\mu$ $_\alpha$ = -3.42 mas yr$^{-1}$ and $\mu$$_\delta$ = 3.82 mas yr$^{-1}$). Furthermore, we estimated an age of 3 $\pm$ 0.2 Gyr and [Fe/H] = 0.5 dex for Pismis 3 (See Figure \ref{Figure5E} bottom panel), which is in excellent agreement with the 3 Gyr computed by \cite{cantat2020painting}. The other physical parameters, such as tangential velocities, vertical-horizontal distance, and core and tidal radius, whose values are consistent with those expected for open clusters, are computed for the cluster and reported in Table \ref{table2}.

In Figure \ref{FigureZ}, we present a sky distribution plot of all the clusters, showing their positions in Galactic coordinates $l$ and $b$, with the colour representing their tangential velocities (V$_T$). The Galactic plane is indicated by the red line at $b$ = 0$^{\circ}$, and the Galactic disk is shown as the shaded area. We observed that the clusters are mostly concentrated near the Galactic plane $b$ $\approx$ 0$^{\circ}$, suggesting that they are part of or close to the Galactic disk, a characteristic observed in other open clusters. This concentration near the Galactic plane indicates that these clusters likely belong to the Milky Way disk.
By the comparison, we can conclude that

a.) While our derived values closely agree with those provided by \cite{cantat2020painting}, they differ from those provided by \cite{kharchenko2013global} and \cite{dias2014proper}.
b.) Unlike in other studies, which study these clusters as a group, we examined each individually in detail in this present analysis.
c.) Eight of the examined clusters have their [Fe/H] values roughly estimated for the first time.

d.) We provided a detailed characterization of the clusters Juchert 12 and Haffner 10 for the first time.

 Finally, in this study, we applied three extinction laws: \cite{schlafly2011measuring},\cite{nataf2015interstellar} and \cite{ruiz2018empirical}. After comparing the results, we adopted and reported results from the two methods that showed the highest agreement. Here are our findings:
 The distances derived using the \cite{ruiz2018empirical} and \cite{nataf2015interstellar} extinction laws demonstrated the highest level of agreement, with absolute differences typically within 0.5 kpc. This similarity likely arises from their shared reliance on robust extinction correction methods. Ruiz-Dern's method integrates precise \textit{Gaia}-based parameters, while Nataf employs advanced bulge calibration techniques. These similarities make both approaches highly suitable for analyzing star clusters in regions with high extinction variations, especially near the Galactic bulge.
 In contrast, distances calculated using the SF11 extinction law consistently deviated, with differences exceeding 1.0 kpc for clusters such as Schuster-MWSC1756 ($b$ = -0.2568), FSR1521 ($b$ = -1.6251), and Saurer 2 ($b$ = -0.9970). These clusters are located in low Galactic latitudes, where extinction is particularly severe. The SF11 method relies on two-dimensional dust maps, which tend to overestimate extinction in dense, heavily reddened regions. As \cite{nataf2015interstellar} pointed out, these maps are less effective at resolving extinction variations in high-density areas, limiting their accuracy for clusters like those examined in this study.

\section{Conclusions}
\label{section7}

We have provided updated measurements for the nine open clusters we studied and, for the first time, a detailed characterization of the cluster Juchert 12. Our derived parameters, where available, agree with those provided by \citep{cantat2020painting}. However, our results conflict with those provided by \cite{kharchenko2013global} and \cite{dias2014proper}. The colour-magnitude diagrams, ages, metallicities, structural parameters, and sky distribution of the clusters indicate that these objects are genuine young, intermediate and old open clusters within the Milky Way disk.

Compared to literature values, our final derived cluster parameters are presented in Table \ref{table2}. We derived cluster reddening, extinction, distance, vertical component, and tangential velocity. Our derived extinctions range from A$_{Ks}$ = 0.07 $\pm$ 0.03  to 0.57 $\pm$ 0.04 , colour excess from E(J - K$_s$) = 0.16 $\pm$ 0.03  to 0.60 $\pm$ 0.03 , distances from D =  2.19 $\pm$ 0.06 kpc to  8.94 $\pm$ 0.06 kpc, and Galactocentric distances from R$_G$ =  7.82 to  15.08 kpc. The Z component ranges from Z -0.09 to 0.34 kpc, and tangential velocities range from V$_T$ = 30.59 to 245.42 km/s. 

We also computed age and metallicity by fitting PARSEC isochrones, finding ages from t = 20 Myr to 5 Gyr and metallicity from [Fe/H] = -0.5 to 0.5 dex. Structural parameters were derived, with core radii ranging from r$_c$ = 0.71 to 5.21', tidal radii from r$_t$ = 3.6' to 12.0', and concentration indices from c = 0.36 to 0.83. These measurements confirm that the nine clusters examined are genuine young, intermediate and old open clusters. Finally, a spectroscopic follow-up is necessary to constrain the metallicity and chemical abundances of the clusters. Additionally, spectroscopic data will help determine the orbital parameters of each cluster.

\begin{table*}[ht]
\centering

\caption{Derived physical parameters for each cluster compared with literature values from \cite{cantat2020painting}, \cite{kharchenko2013global}, and \cite{dias2014proper}.}
\label{table2}
\resizebox{\textwidth}{!}{%
\begin{tabular}{@{}llllllllllllllllll@{}}
\toprule
Name & $\mu_{\alpha*}$ [mas yr$^{-1}$] & $\mu_{\delta}$ [mas yr$^{-1}$] & $A_G$ & D$_{Gaia}$[kpc] & D$_{avg}$ [kpc] & $R_G$ [kpc] & $Z$ [kpc] & E(B-V) & Age [Gyr] & $r_c$ [pc] & $r_t$ [pc] & $c$ & V$_{T\alpha*}$ [km/s] & V$_{T\delta}$ [km/s] & $V_T$ [km/s] & [Fe/H] [dex] \\ \midrule
BH118 & $-5.493\pm0.209$ & $1.519\pm0.280$ & 1.39 & 7.10 & 8.34$\pm$0.03 & 9.29 & 0.34 & 0.53$\pm$0.05& 3$\pm$0.5 & 1.78 & 11.89 & 0.83 & $-225.42$ & $62.31$ & $233.88$ & 0.5 \\
BH\_144 & $-5.338\pm0.095$ & $-0.775\pm0.780$ & 2.16 & 6.60 & 8.81$\pm$0.10 & 7.82&  -0.49& 0.81$\pm$0.06& 2$\pm$0.2 & 1.9 & 11.88 & 0.78 & $-242.76$ & $-36.07$ & $245.42$ & 0.5 \\
Schuster-MWSC1756 & $5.244\pm0.199$ & $3.167\pm0.231$ & 9.25 & - & 7.03$\pm$0.05 & 9.73& -0.03 &3.47$\pm$0.20 & 700$\pm$0.2 Myrs & 1.4 & 7.04 & 0.70 & $167.69$ & $100.67$ & $195.59$ & 0.0 \\
Saurer3 & $-5.043\pm0.201$ & $2.232\pm0.909$ & 2.20 & 8.70 & 7.23$\pm$0.04 & 9.41 & 0.38& 0.83$\pm$0.06 & 2$\pm$0.2 & 1.6 & 9.6 & 0.77 & $-221.85$ & $108.58$ & $246.99$ & -0.5 \\
FSR1521 & $-4.827\pm0.050$ & $3.317\pm0.359$ & 8.87 & - & 3.15$\pm$0.05 & 8.23& -0.09 &3.33$\pm$0.20 & 1$\pm$0.2 & 2.5 & 8.9 & 0.55 & $-77.82$ & $68.41$ & $103.61$ & 0.0 \\
Saurer2 & $-2.362\pm0.249$ & $2.729\pm0.342$ & 5.02 & - & 5.49$\pm$0.05 & 10.77& -0.10&1.89$\pm$0.11 & 3$\pm$0.2 & 1.4 & 7.8 & 0.75 & $-71.17$ & $95.14$ & $118.82$ & 0.0 \\
Haffner10-FSR1231 & $-1.311\pm0.941$ & $1.436\pm0.772$ & 1.76 & 4.00 & 3.62$\pm$0.03 & 10.86 & 0.06 & 0.66$\pm$0.05& 20$\pm$0.2 Myrs & 1.2 & 6.8 & 0.74 & $-19.44$ & $23.62$ & $30.59$ & 0.5 \\
Juchert12 & $-0.872\pm0.026$ & $1.539\pm0.051$ & 2.23 & 8.70 & 8.94$\pm$0.06 & 15.08& -0.64 & 0.84$\pm$0.06& 2$\pm$0.2 & 2.4 & 14.5 & 0.79 & $-30.72$ & $66.34$ & $73.11$ & -0.5 \\
Pismis3 & $-4.755\pm0.377$ & $6.655\pm0.011$ & 3.75 & 1.10 & 2.19$\pm$0.06 & 8.92 & 0.02 &1.41$\pm$0.09 & 3$\pm$0.2 & 3.18 & 7.3 & 0.36 & $-46.43$ & $64.98$ & $79.86$ & 0.5 \\
\midrule
\multicolumn{17}{c}{\cite{cantat2020painting} measurement} \\ 
\midrule
BH118 & -4.406 & 1.450 & -- & -- & 8.8 & 9.6 & 0.36 & -- & -- & -- & -- & -- & -- & -- & -- & -- \\
BH\_144 & -5.111 & -0.375 & -- & -- & 9.6 & 8.3 & -0.532 & -- & -- & -- & -- & -- & -- & -- & -- & -- \\
Schuster-MWSC1756 & 5.102 & 3.760 & -- & -- & -- & -- & -- & -- & -- & -- & -- & -- & -- & -- & -- & -- \\
Saurer3 & -- & -- & -- & -- & -- & -- & -- & -- & -- & -- & -- & -- & -- & -- & -- & -- \\
FSR1521 & -4.886 & 4.467 & -- & -- & 4.933 & 8.886 & -0.139 & -- & -- & -- & -- & -- & -- & -- & -- & -- \\
Saurer2 & -2.969 & 4.567 & -- & -- & 6.4 & -- & -0.127 & -- & -- & -- & -- & -- & -- & -- & -- & -- \\
Haffner10-FSR1231 & -1.144 & 1.557 & -- & -- & 3.4 & 10.8 & 0.06 & -- & -- & -- & -- & -- & -- & -- & -- & -- \\
Juchert12 & -- & -- & -- & -- & -- & -- & -- & -- & -- & -- & -- & -- & -- & -- & -- & -- \\
Pismis3 & -4.809 & 6.645 & -- & -- & 2.3 & 9.1 & 0.02 & -- & -- & -- & -- & -- & -- & -- & -- & -- \\
\midrule
\multicolumn{17}{c}{\cite{kharchenko2013global} measurement} \\ 
\midrule
BH118 & -- & -- & -- & -- & -- & -- & -- & -- & -- & -- & -- & -- & -- & -- & -- & -- \\
BH\_144 & -13.19 & 7.74 & -- & -- & 7.2 & -- & -- & 0.63 & 1.0 & 0.83 & 12.88 & -- & -- & -- & -- & -0.5 \\
Schuster-MWSC1756 & -9.41 & 3.06 & -- & -- & 1.8 & -- & -- & 1.08 & 6.3 Myrs & 0.26 & 5.41 & -- & -- & -- & -- & -- \\
Saurer3 & -8.34 & 4.40 & -- & -- & 7.1 & -- & -- & 0.78 & 1.9 & 9.18 & 20.46 & -- & -- & -- & -- & -- \\
FSR1521 & -8.92 & 6.10 & -- & -- & 1.5 & -- & -- & 0.38 & 5.1 Myrs & 0.84 & 6.69 & -- & -- & -- & -- & -- \\
Saurer2 & -6.62 & 8.13 & -- & -- & 6.0 & -- & -- & 1.25 & 1.7 & 1.40 & 15.56 & -- & -- & -- & -- & -- \\
Haffner10-FSR1231 & -- & -- & -- & -- & -- & -- & -- & -- & -- & -- & -- & -- & -- & -- & -- & -- \\
Juchert12 & -- & -- & -- & -- & -- & -- & -- & -- & -- & -- & -- & -- & -- & -- & -- & -- \\
Pismis3 & -- & -- & -- & -- & -- & -- & -- & -- & -- & -- & -- & -- & -- & -- & -- & -- \\
\midrule
\multicolumn{17}{c}{\cite{dias2014proper} measurement} \\ 
\midrule
BH118 & -- & -- & -- & -- & -- & -- & -- & -- & -- & -- & -- & -- & -- & -- & -- & -- \\
BH\_144 & -4.73 & -3.36 & -- & -- & -- & -- & -- & -- & -- & -- & -- & -- & -- & -- & -- & -- \\
Schuster-MWSC1756 & -- & -- & -- & -- & -- & -- & -- & -- & -- & -- & -- & -- & -- & -- & -- & -- \\
Saurer3 & -6.73 & -1.62 & -- & -- & -- & -- & -- & -- & -- & -- & -- & -- & -- & -- & -- & -- \\
FSR1521 & -- & -- & -- & -- & -- & -- & -- & -- & -- & -- & -- & -- & -- & -- & -- & -- \\
Saurer2 & -- & -- & -- & -- & -- & -- & -- & -- & -- & -- & -- & -- & -- & -- & -- & -- \\
Haffner10-FSR1231 & -1.04 & -0.36 & -- & -- & -- & -- & -- & -- & -- & -- & -- & -- & -- & -- & -- & -- \\
Juchert12 & -3.46 & 0.52 & -- & -- & -- & -- & -- & -- & -- & -- & -- & -- & -- & -- & -- & -- \\
Pismis3 & -3.42 & 3.82 & -- & -- & -- & -- & -- & -- & -- & -- & -- & -- & -- & -- & -- & -- \\
\bottomrule
\end{tabular}%
}

\end{table*}

\begin{acknowledgements} 
 We gratefully acknowledge the use of data from the ESO Public Survey program IDs 179.B-2002 and 198.B-2004 taken with the VISTA telescope and data products from the Cambridge Astronomical Survey Unit. This work has made use of data from the European Space Agency (ESA) mission  \textit{Gaia} (https://www.cosmos.esa.int/gaia), processed by the  \textit{Gaia} Data Processing and Analysis Consortium (DPAC,\url{ https: //www.cosmos.esa.int/web/gaia/dpac/consortium)}. Funding for the DPAC has been provided by national institutions, in particular, the institutions participating in the  \textit{Gaia} Multilateral Agreement.  This publication makes use of data products from the Two Micron All Sky Survey, which is a joint project of the University of Massachusetts and the Infrared Processing and Analysis Center/California Institute of Technology, funded by the National Aeronautics and Space Administration and the National Science Foundation.
  We would also like to thank the reviewer for the constructive feedback, which has improved the quality of our manuscript and strengthened our discussion.
 C.O.Obasi gratefully acknowledges the grants support provided by the Joint Committee ESO-Government of Chile under the agreement 2023 ORP 062/2023.
 J.G.F-T gratefully acknowledges the grants support provided by ANID Fondecyt Iniciaci\'on No. 11220340, ANID Fondecyt Postdoc No. 3230001 (Sponsoring researcher) from the Joint Committee ESO-Government of Chile under the agreement 2021 ORP 023/2021 and 2023 ORP 062/2023.
 D.M. gratefully acknowledges support from the Center for Astrophysics and Associated Technologies CATA by the ANID BASAL projects ACE210002 and FB210003, by Fondecyt Project No. 1220724. M.G. gratefully acknowledges support from Fondecyt through grant 1240755

\end{acknowledgements}

\bibliographystyle{aa}
\bibliography{references}
\begin{appendix}
\onecolumn
\section{Additional figures}
\begin{figure*}[t]
\begin{center}
\includegraphics[height = 4.5cm]{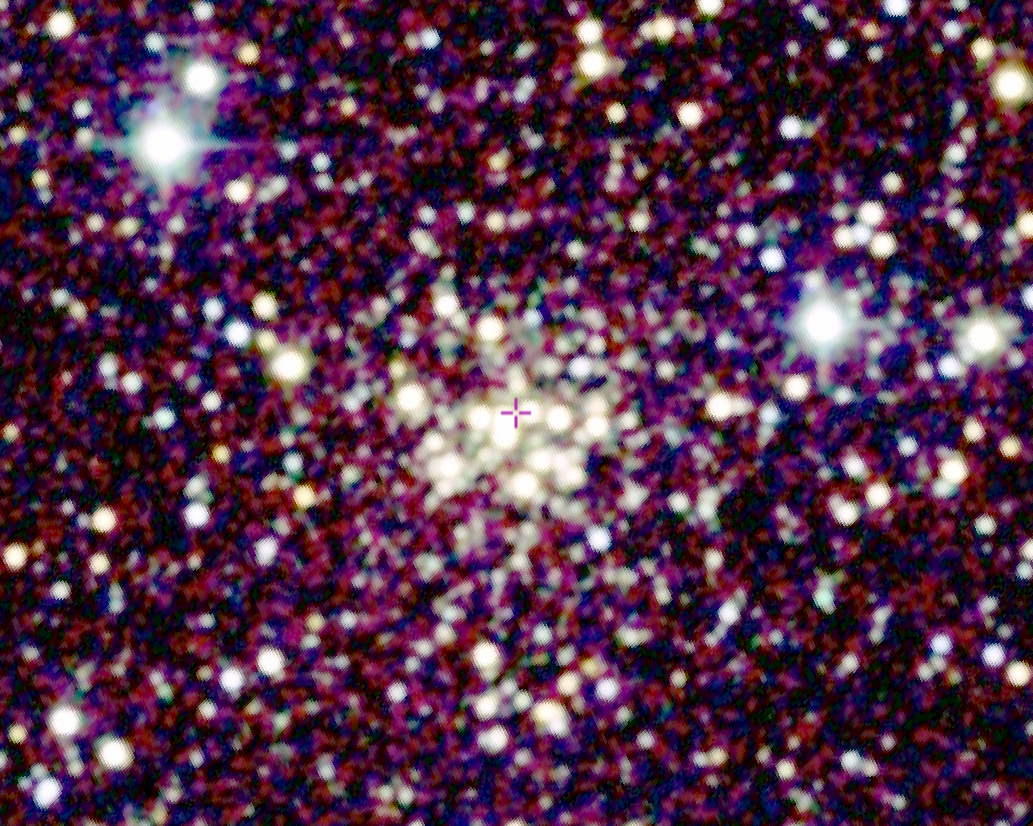}
\includegraphics[height = 4.5cm]{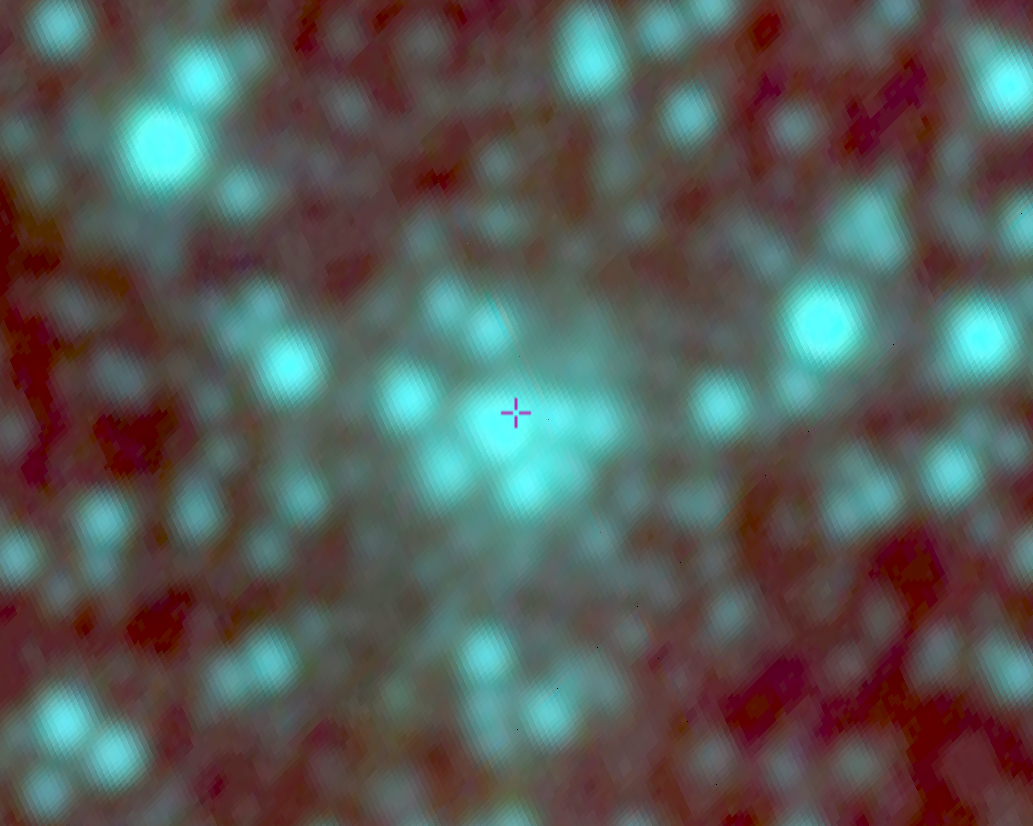}
\includegraphics[height = 4.5cm]{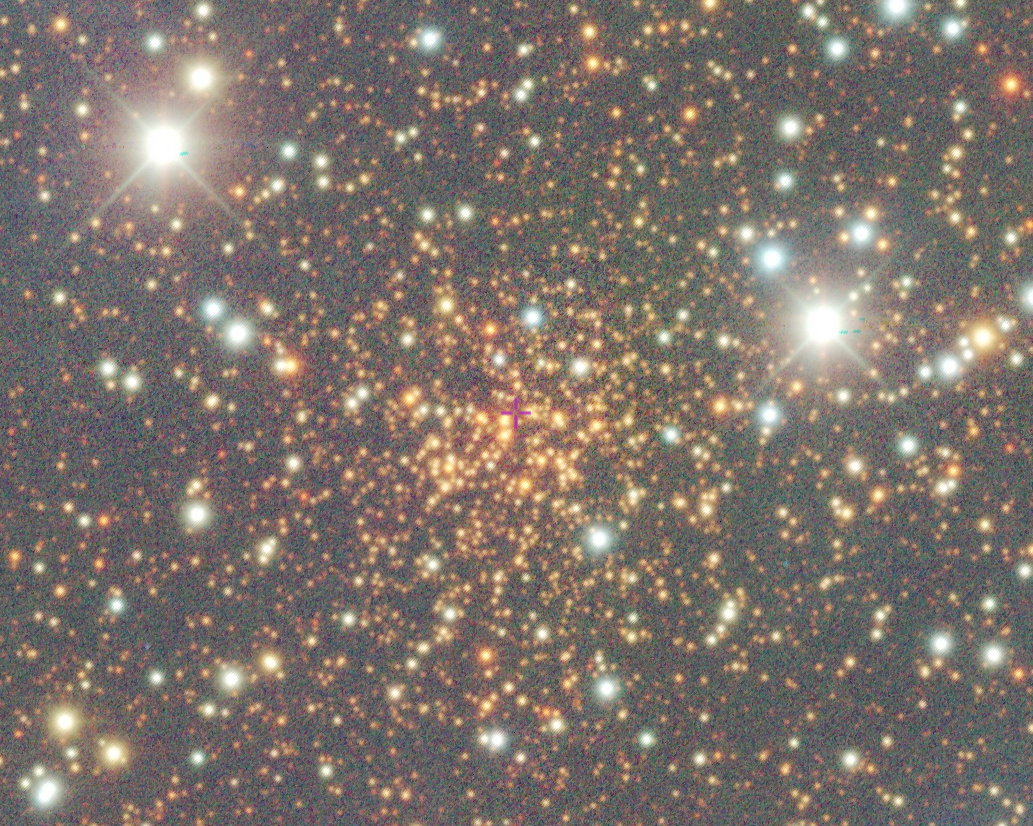}\\
\includegraphics[height = 4.5cm]{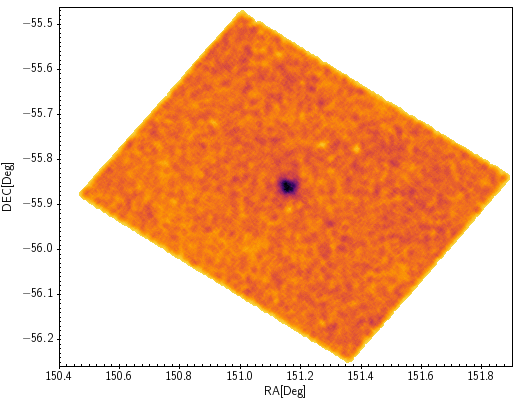}
\includegraphics[height = 5cm]{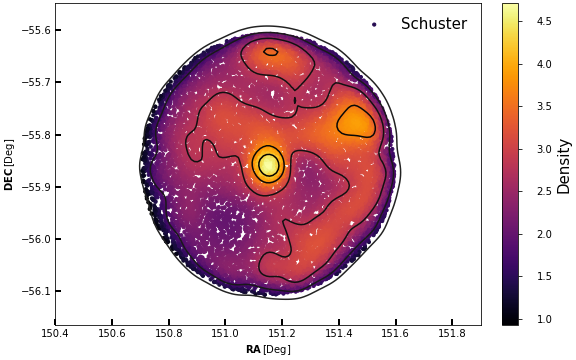}\\
\includegraphics[height = 4.5cm]{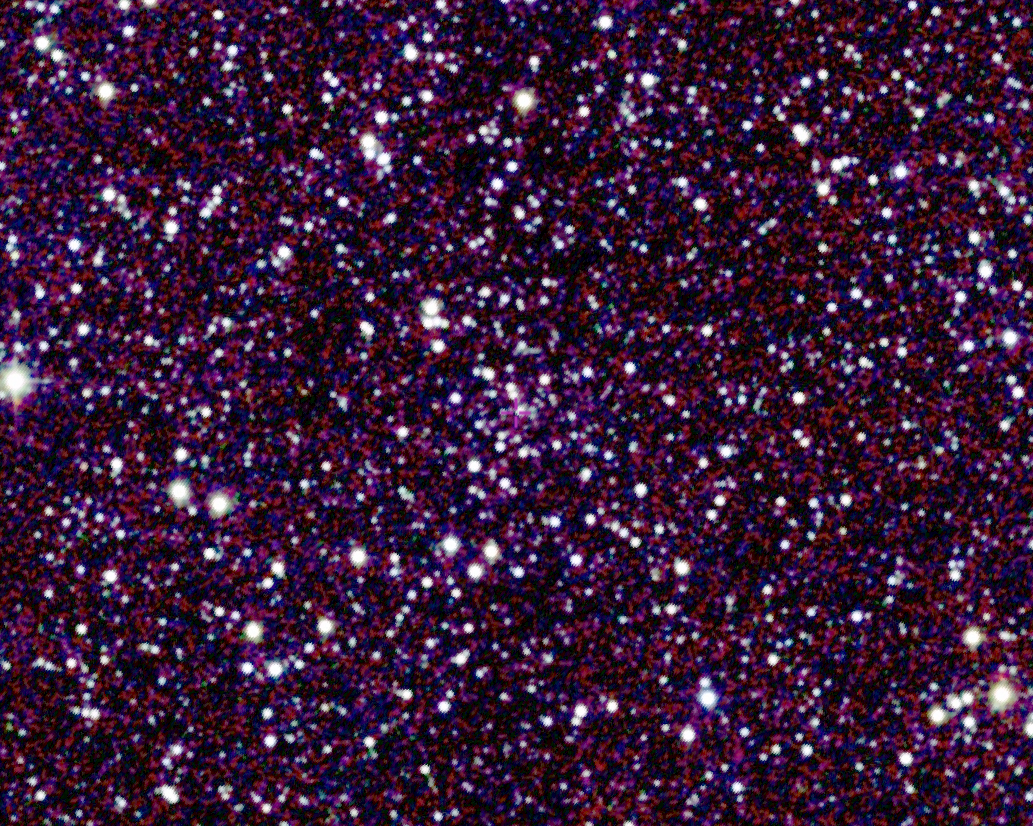}
\includegraphics[height = 4.5cm]{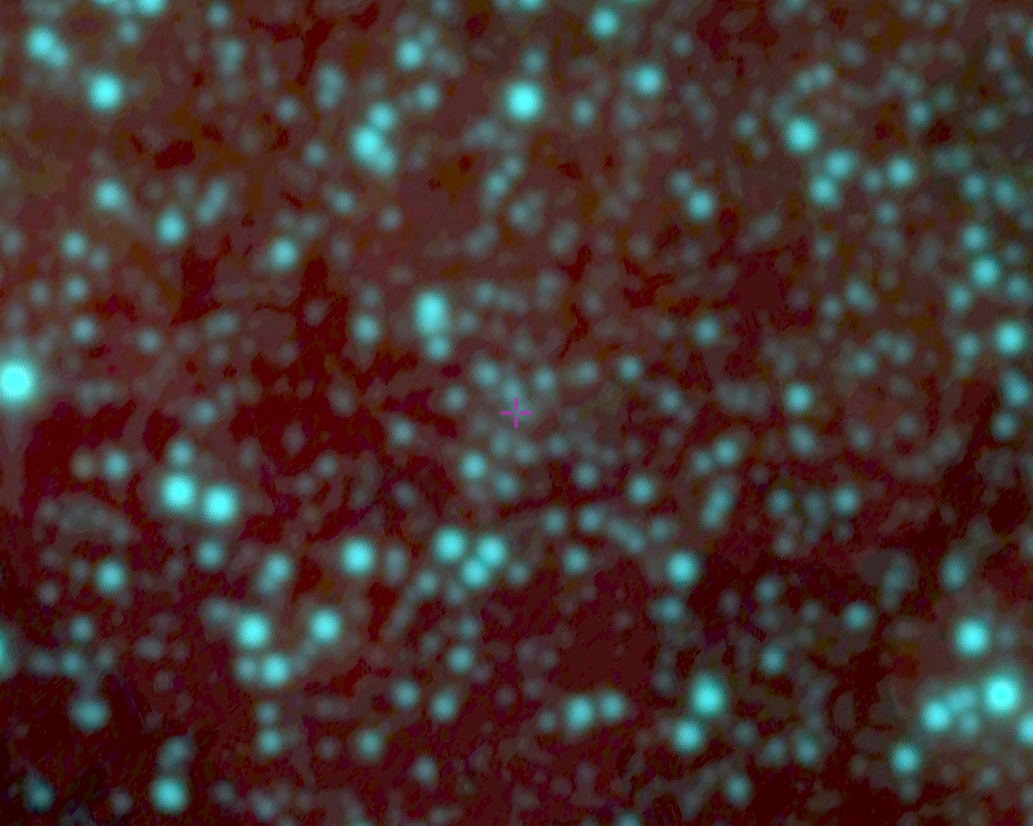}
\includegraphics[height = 4.5cm]{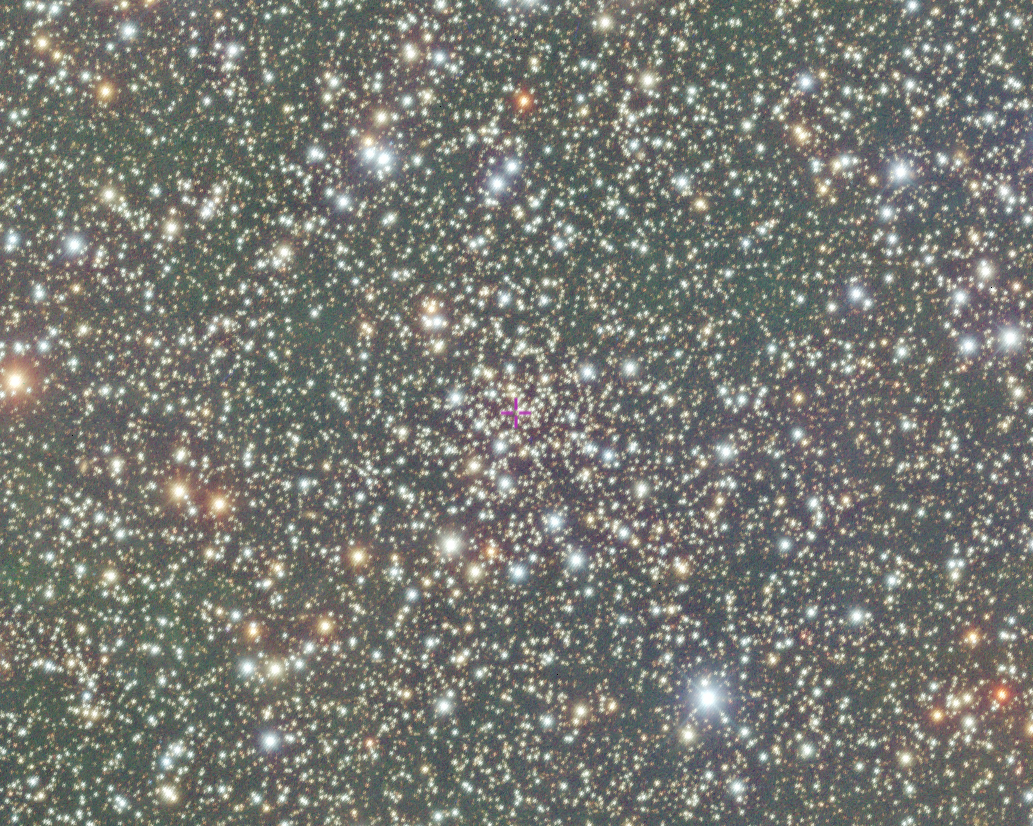}\\
\includegraphics[height = 5cm]{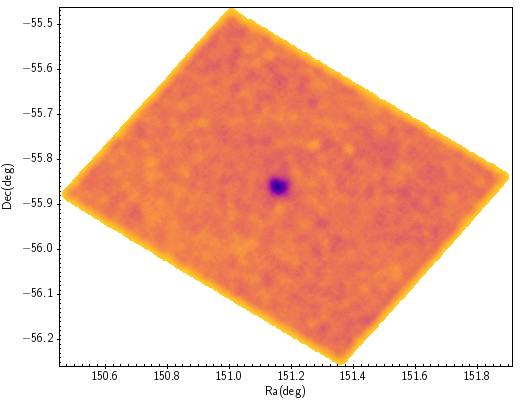}
\includegraphics[height = 5cm]{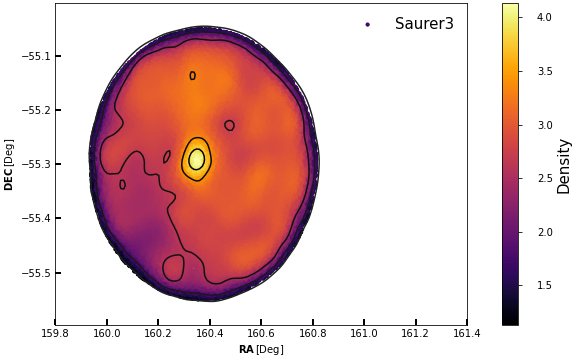}

\caption{Same as Fig. 2, but for the clusters Schuster and Sauer 3.}
\label{Figure2B}
\end{center}
\end{figure*}

 \begin{figure*}[t]
\begin{center}
\includegraphics[height = 4.5cm]{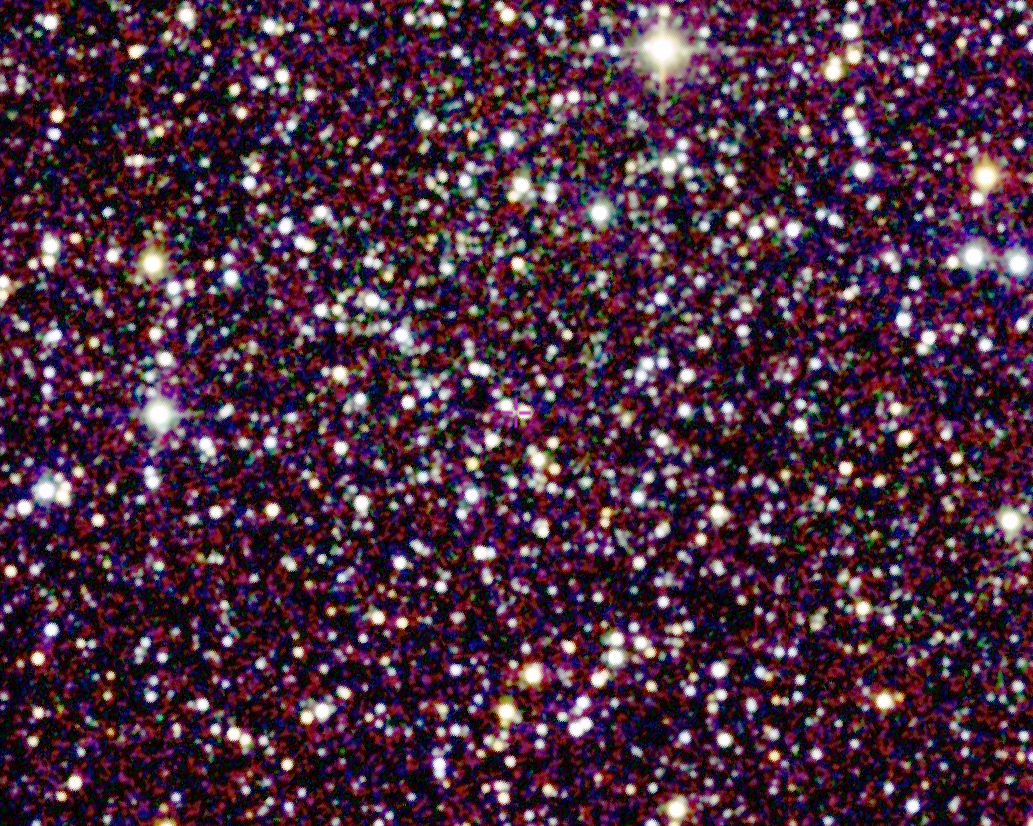}
\includegraphics[height = 4.5cm]{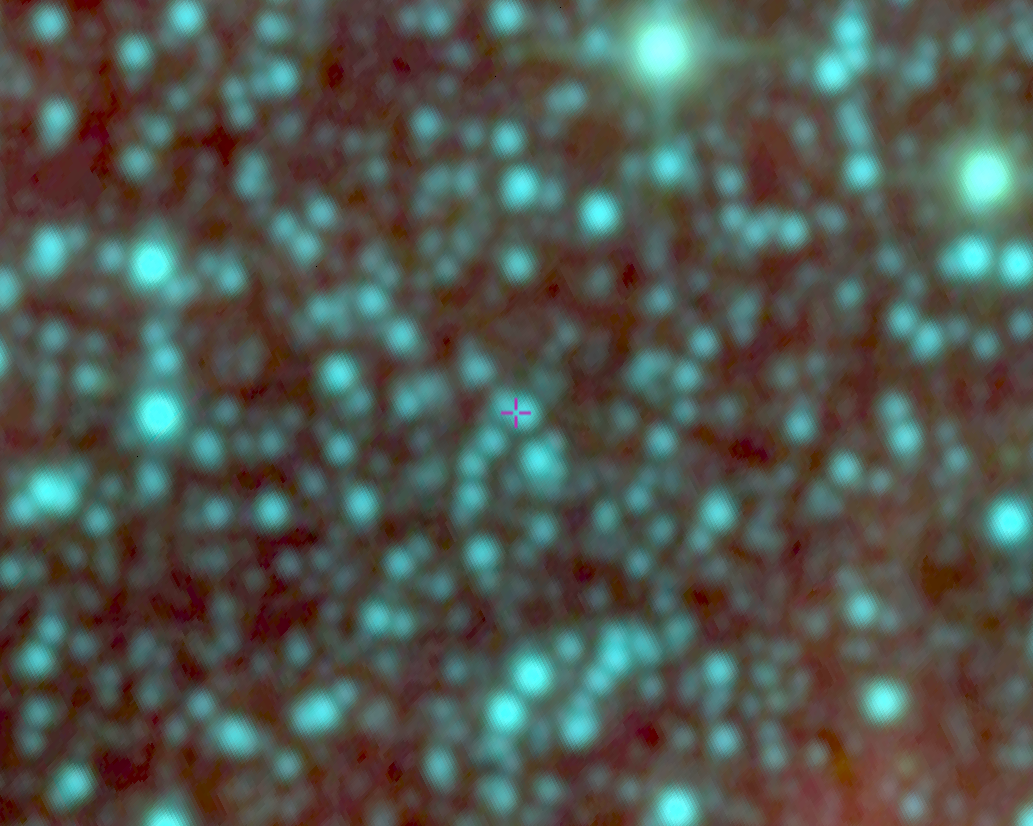}
\includegraphics[height = 4.5cm]{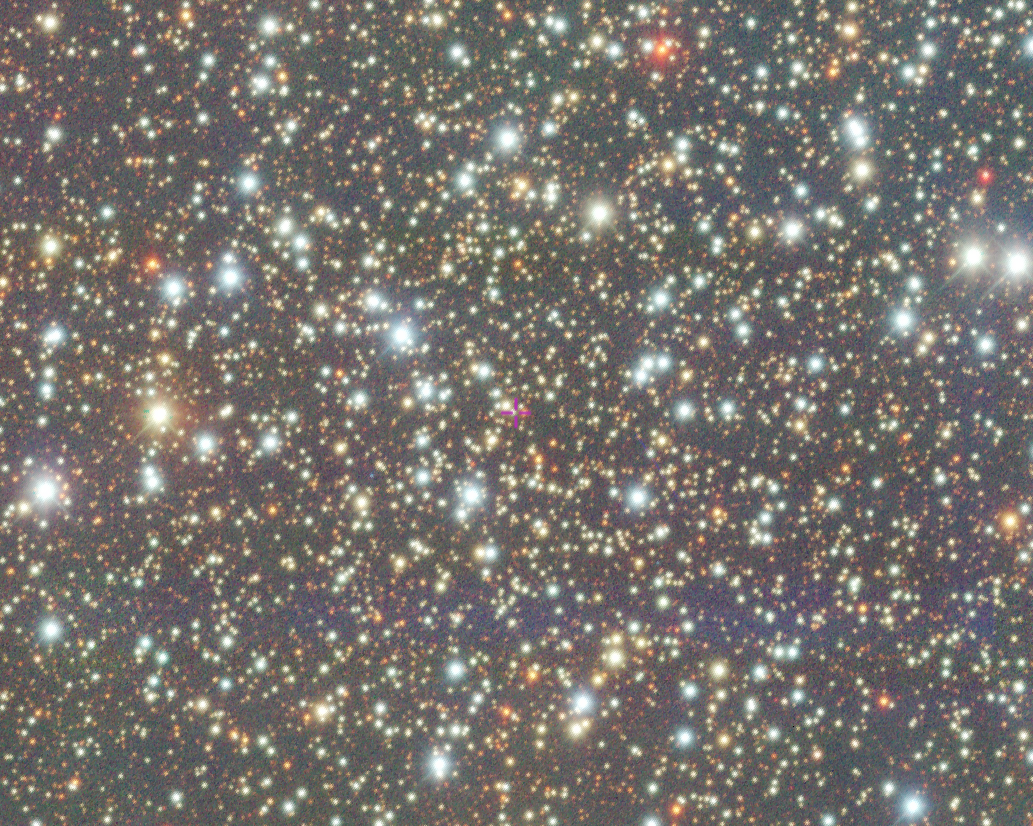}\\
\includegraphics[height = 5cm]{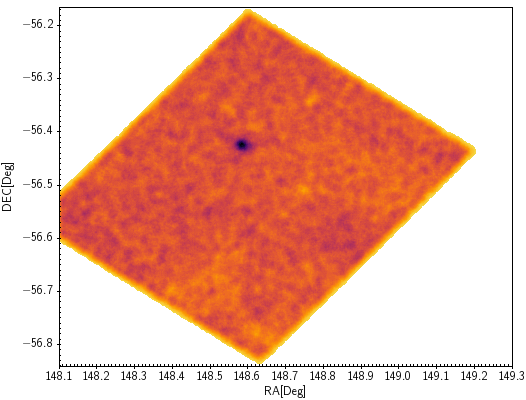}
\includegraphics[height = 5cm]{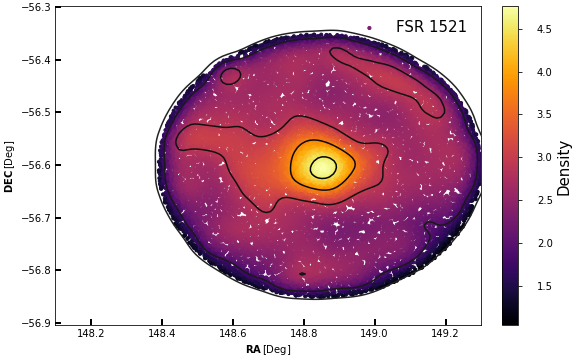}\\
\includegraphics[height = 4.5cm]{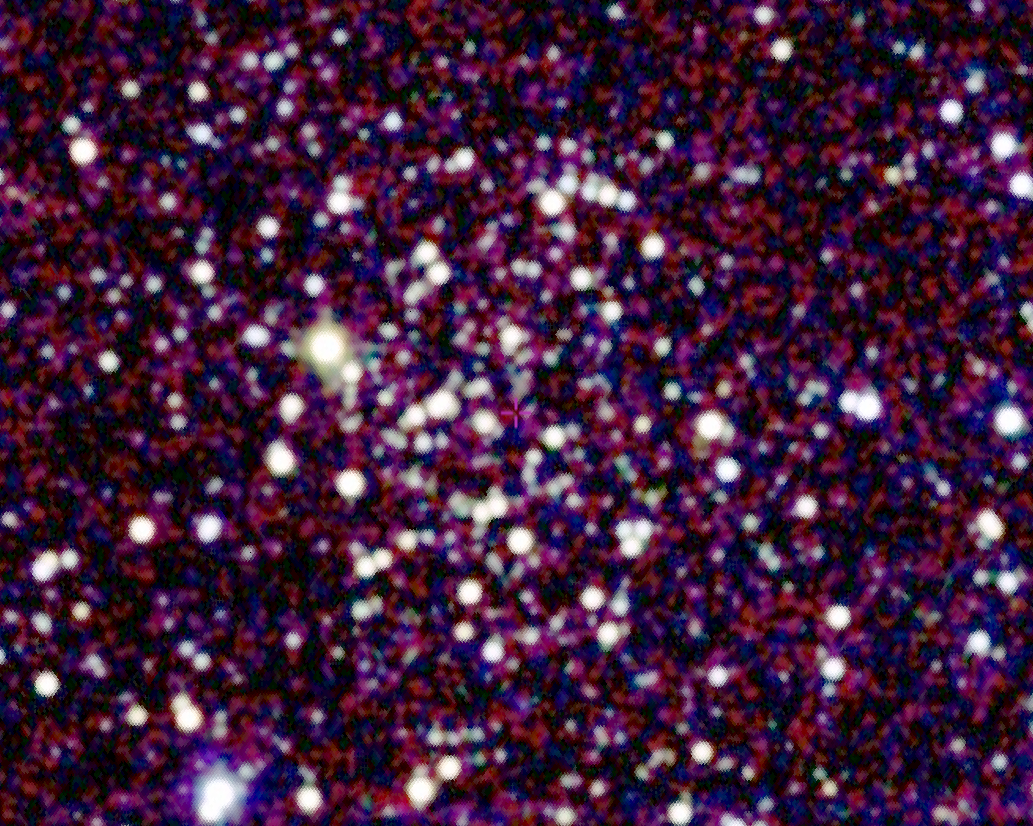}
\includegraphics[height = 4.5cm]{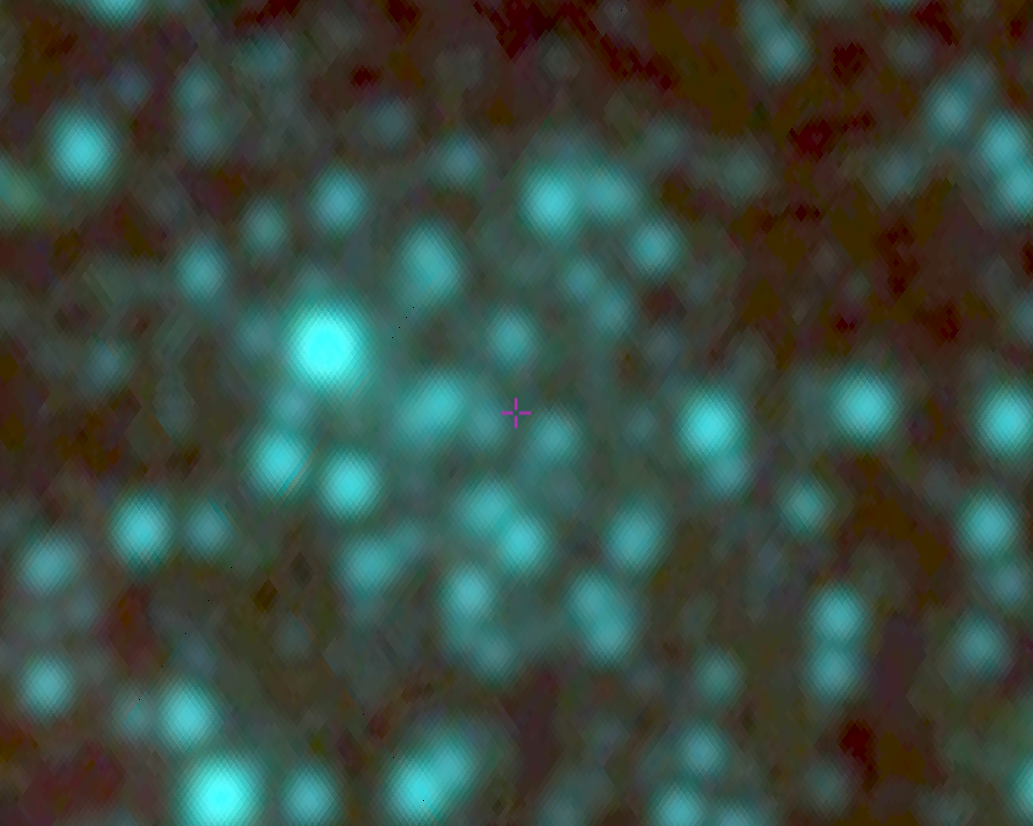}
\includegraphics[height = 4.5cm]{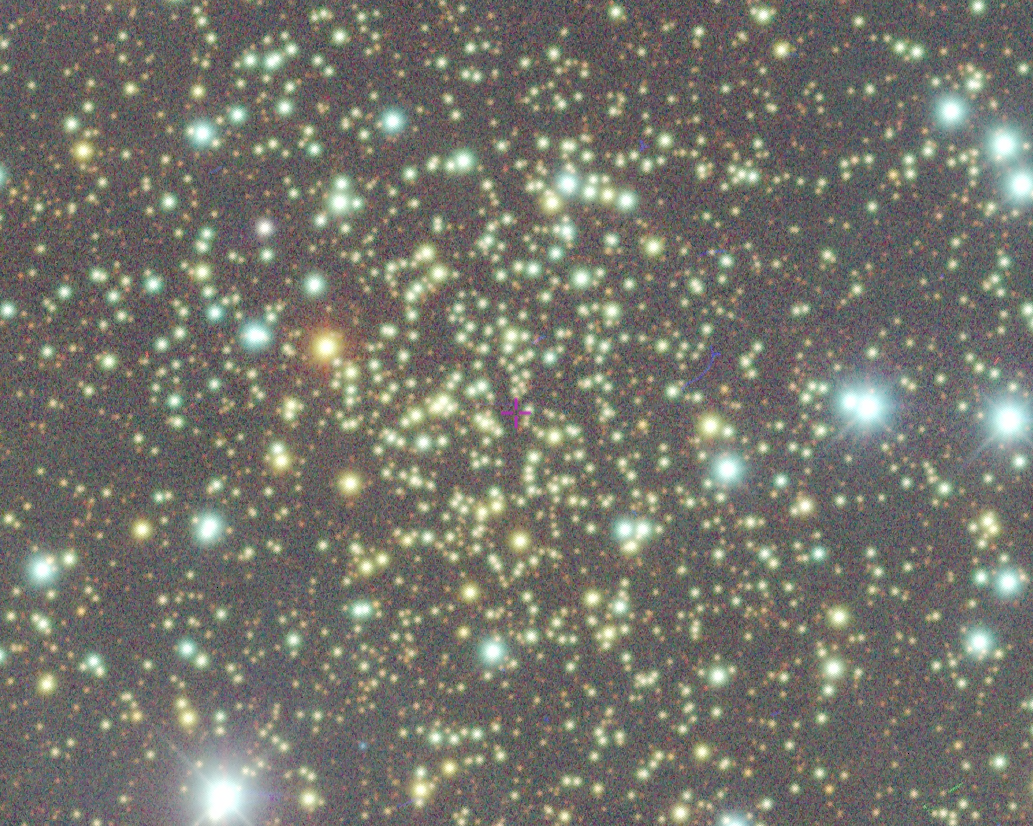}\\
\includegraphics[height = 5cm]{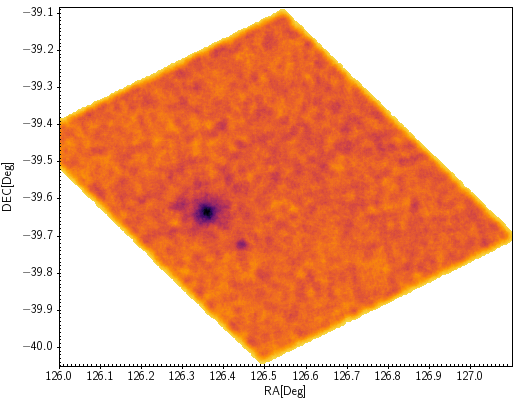}
\includegraphics[height = 5cm]{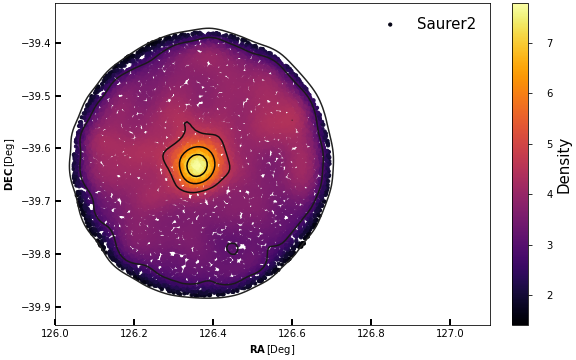}

\caption{Same as Fig. 2, but for the clusters FSR 1521 and Sauer 2.}
\label{Figure2C}
\end{center}
\end{figure*}

\begin{figure*}[t]
\begin{center}
\includegraphics[height = 4.5cm]{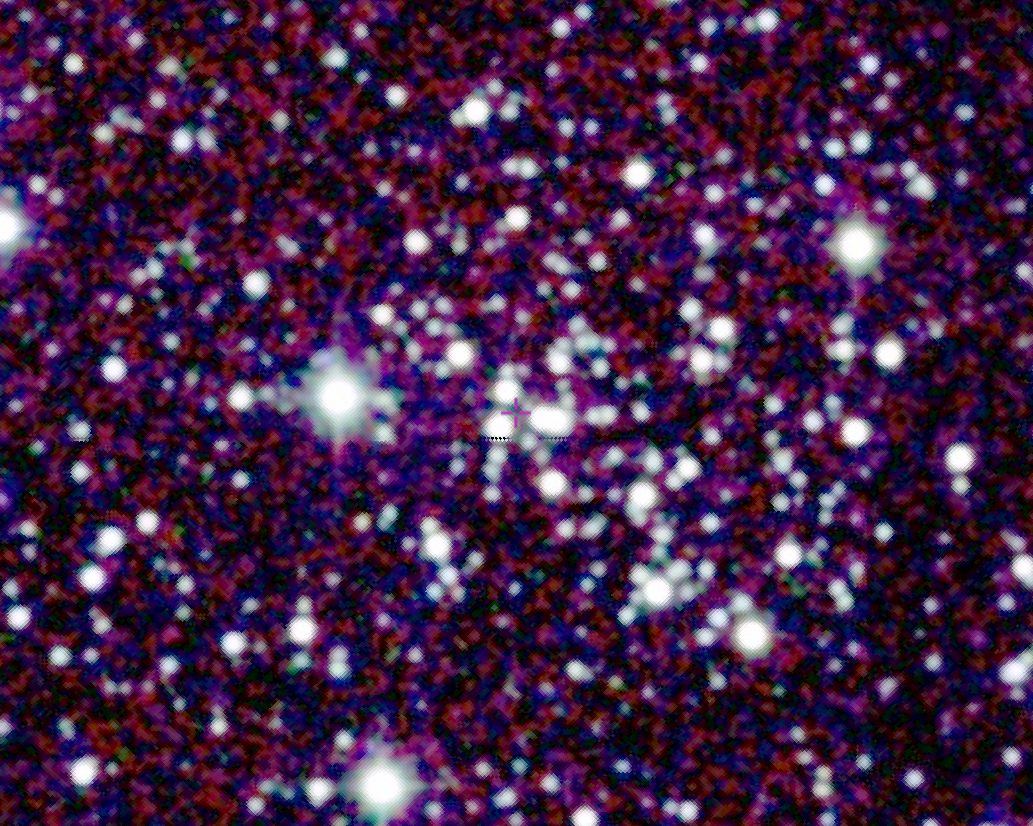}
\includegraphics[height = 4.5cm]{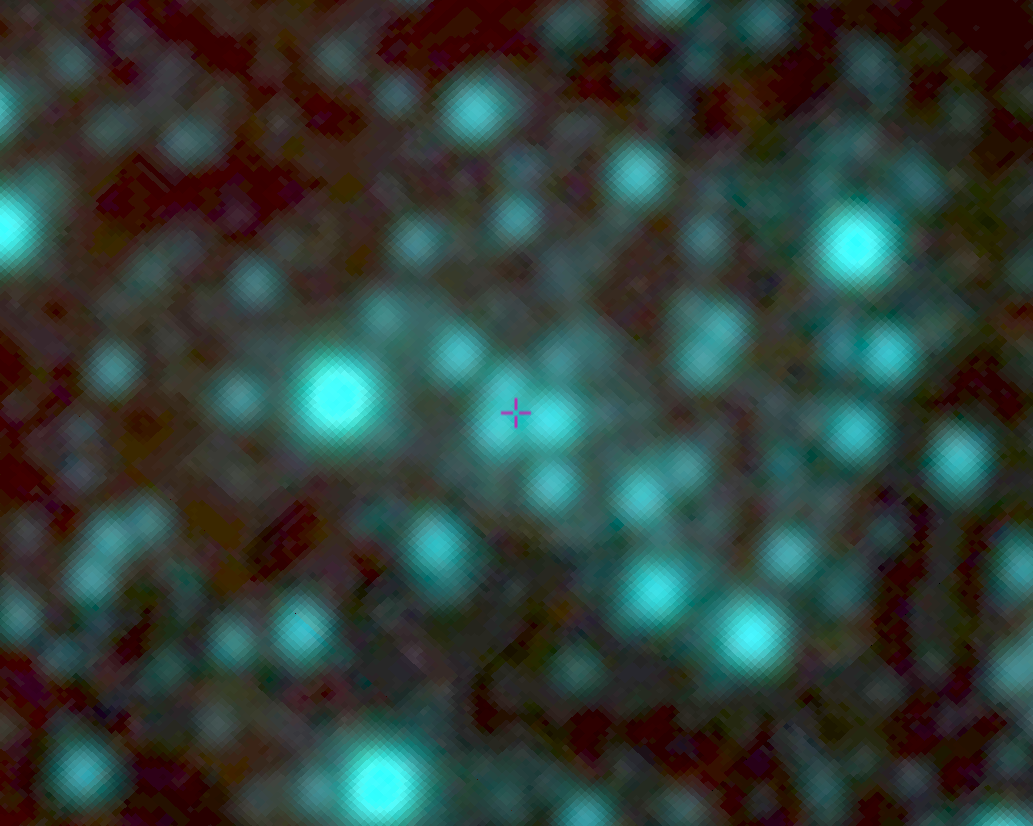}
\includegraphics[height = 4.5cm]{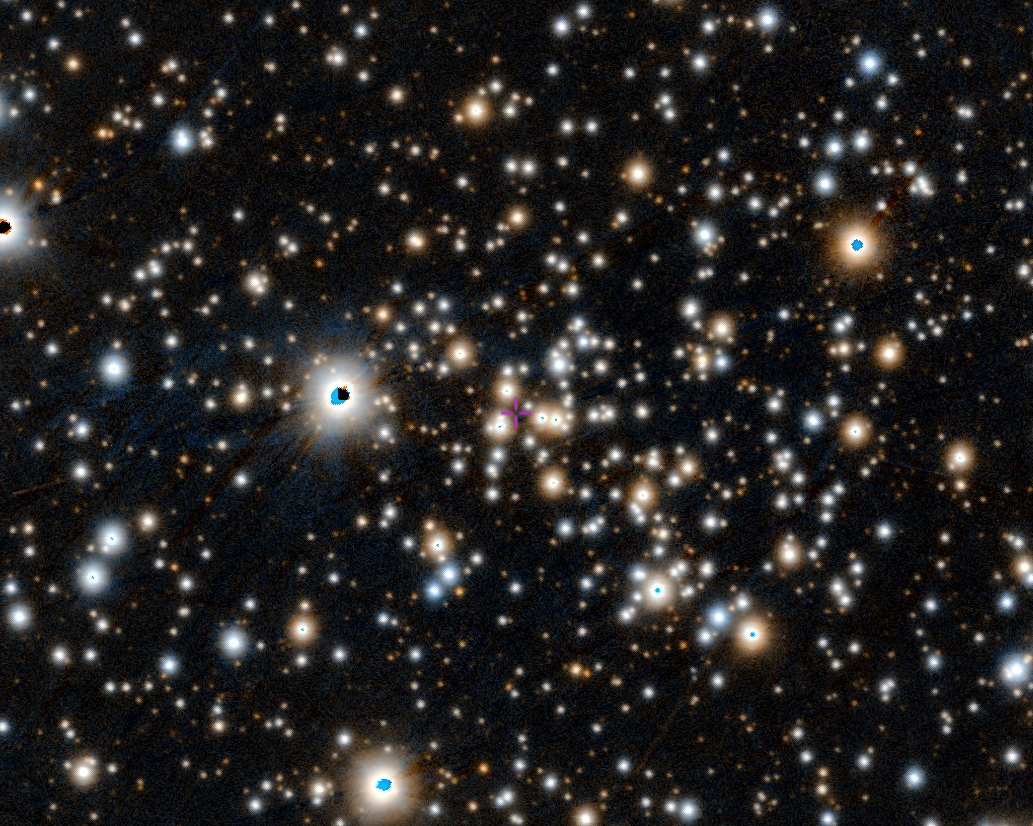}\\
\includegraphics[height = 5cm]{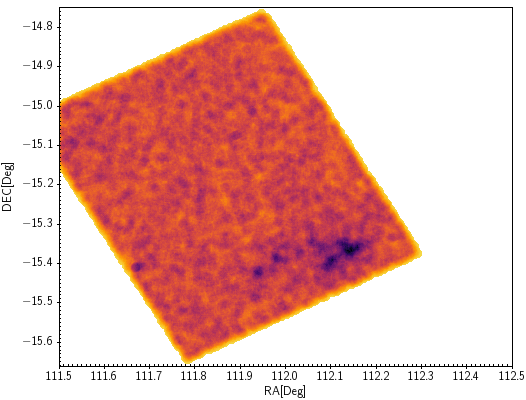}
\includegraphics[height = 5cm]{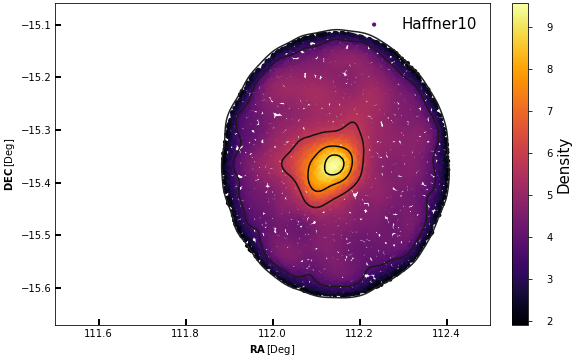}\\
\includegraphics[height = 4.5cm]{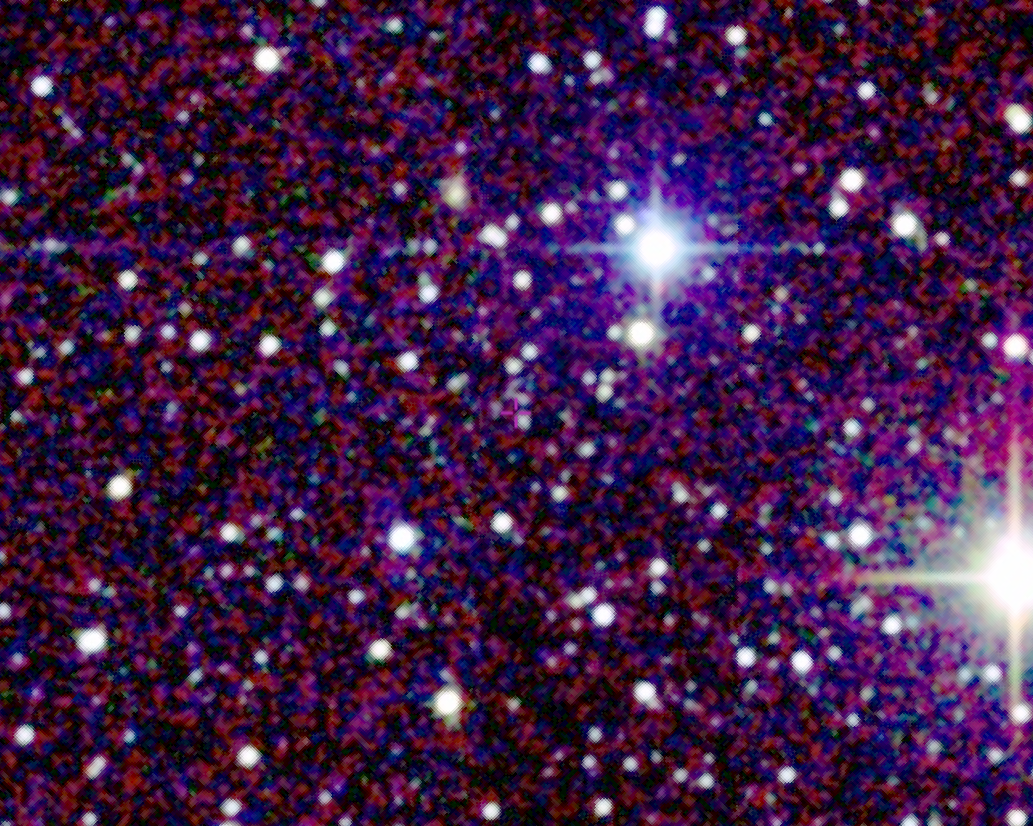}
\includegraphics[height = 4.5cm]{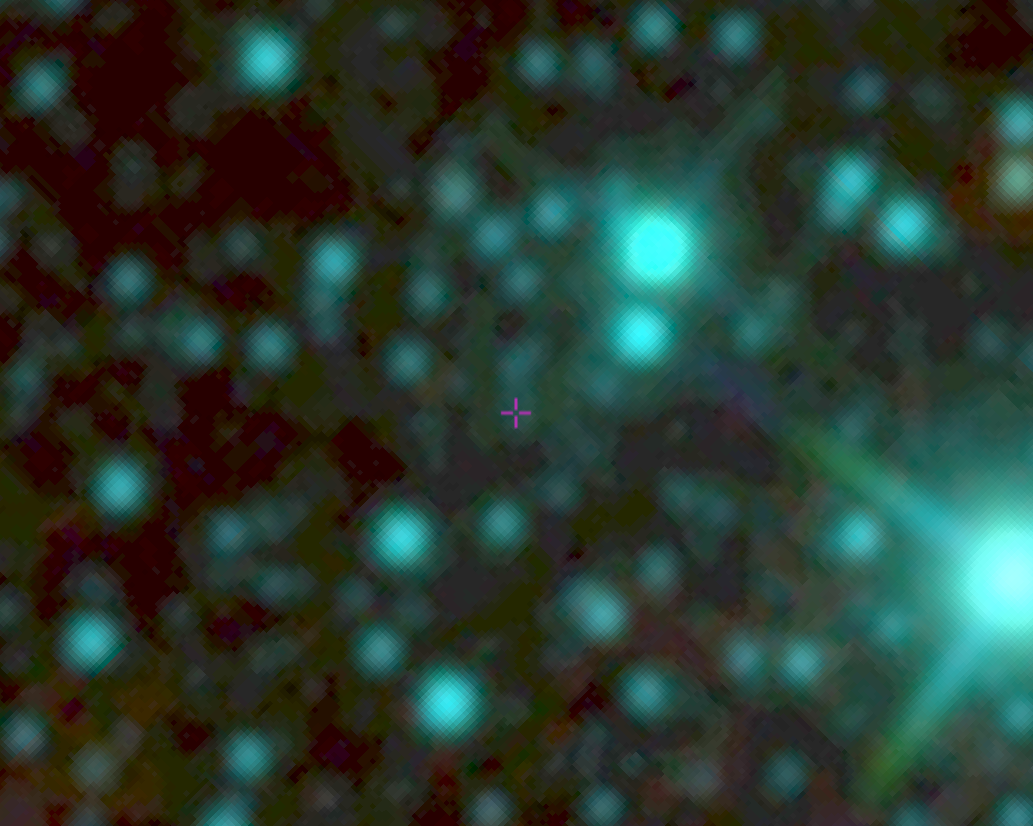}
\includegraphics[height = 4.5cm]{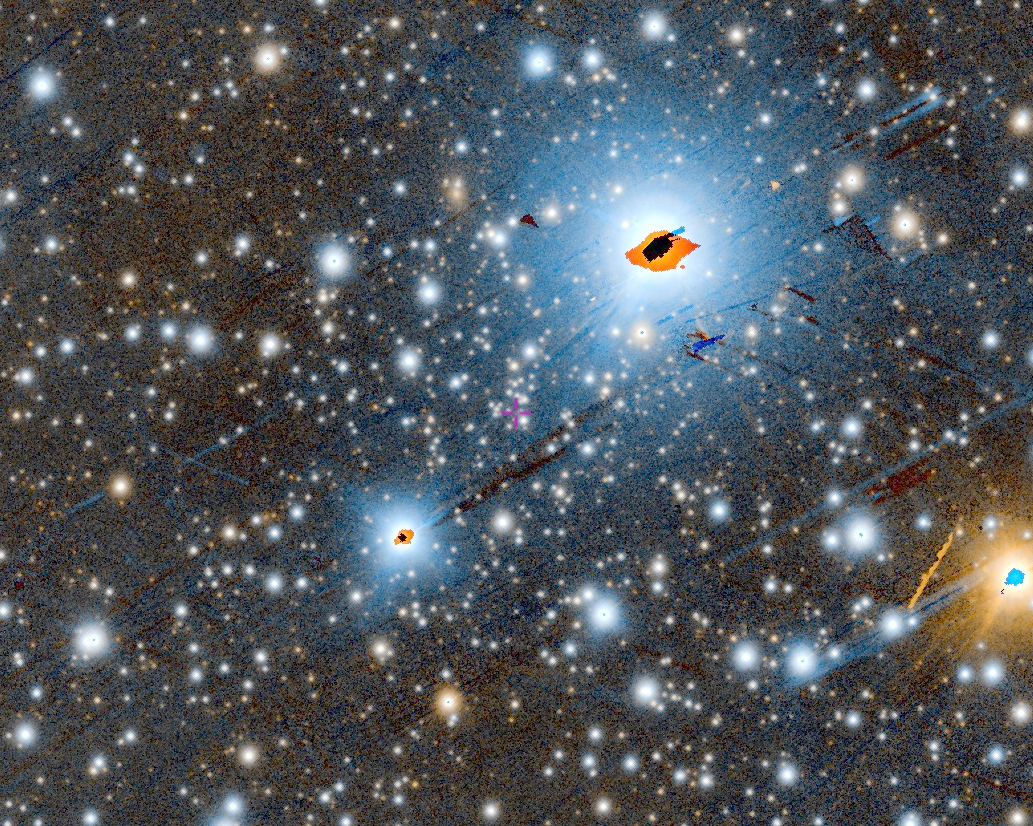}\\

\includegraphics[height = 5cm]{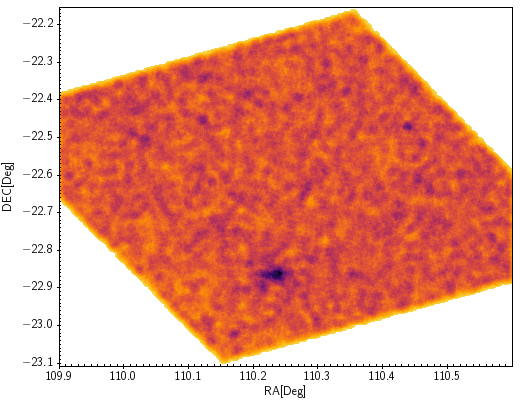}
\includegraphics[height = 5cm]{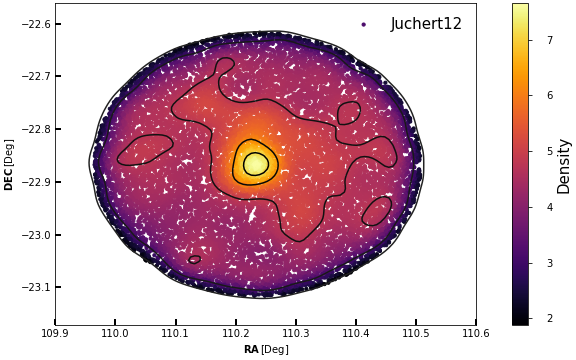}
\caption{Same as Fig. 2, but for the clusters Haffner10 and Juchert 12, however, the right image in the first and third panels is from Pan-STARRS.}
\label{Figure2D}
\end{center}
\end{figure*}

\begin{figure*}[t]
\begin{center}

\includegraphics[height=4.5cm]{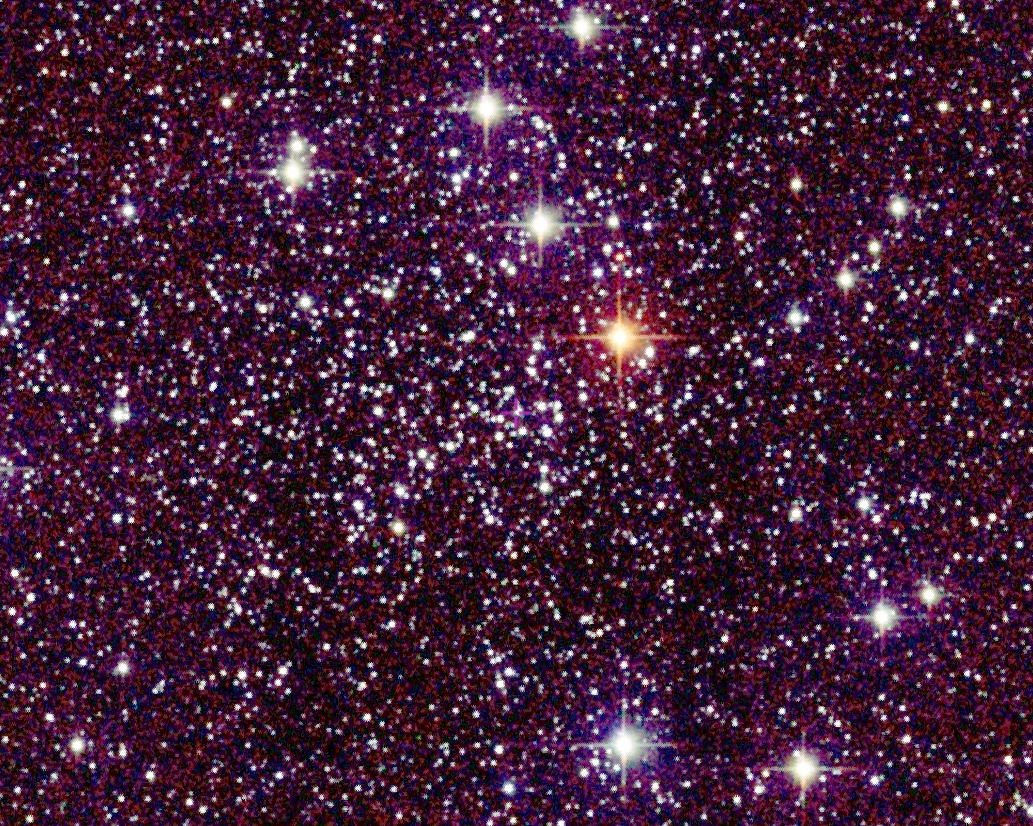}
\includegraphics[height=4.5cm]{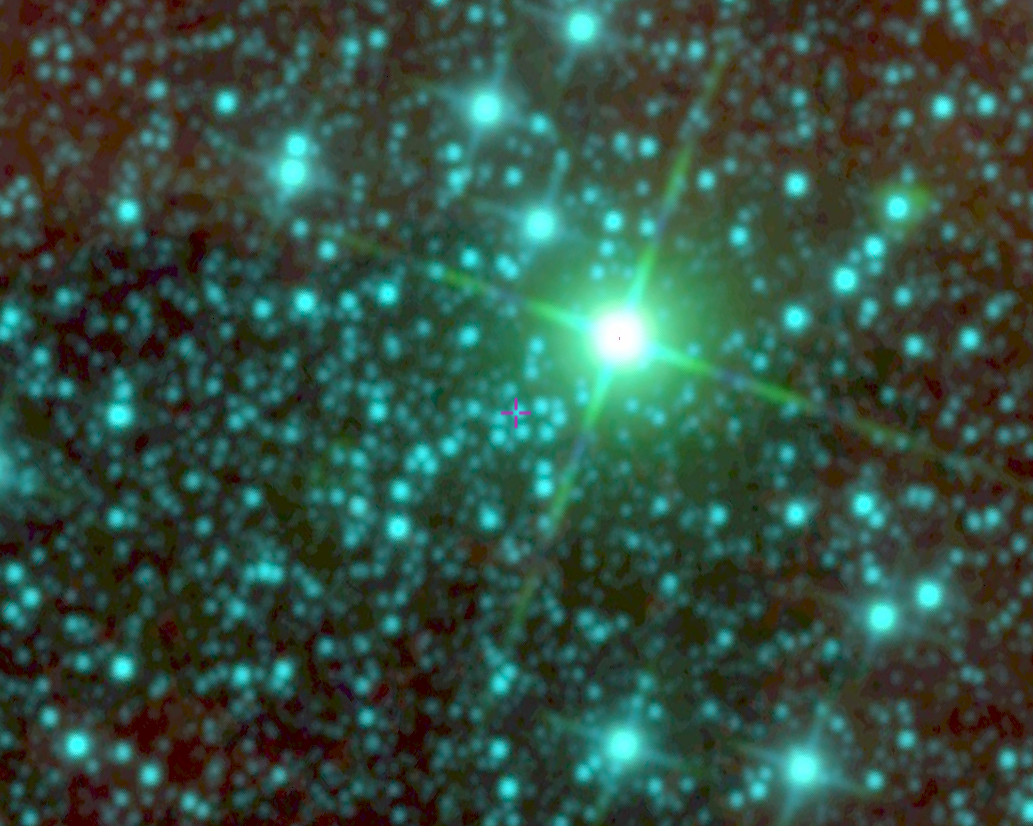}
\includegraphics[height=4.5cm]{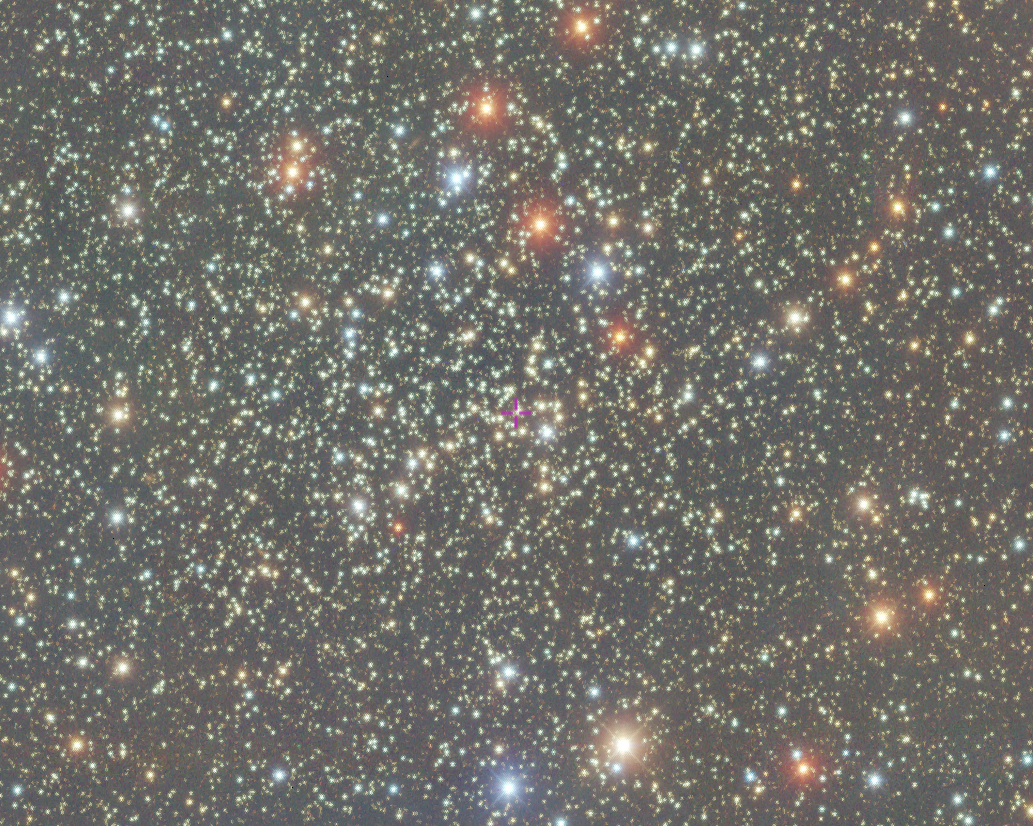}\\
\includegraphics[height=5cm]{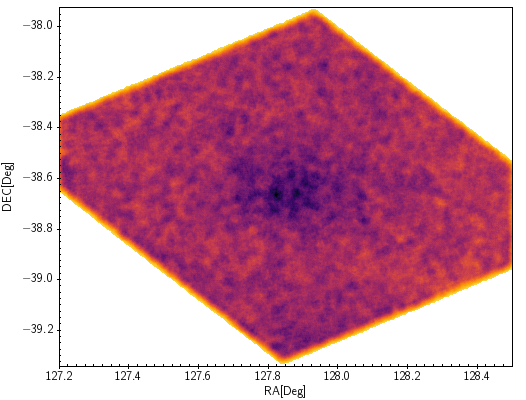}
\includegraphics[height=5cm]{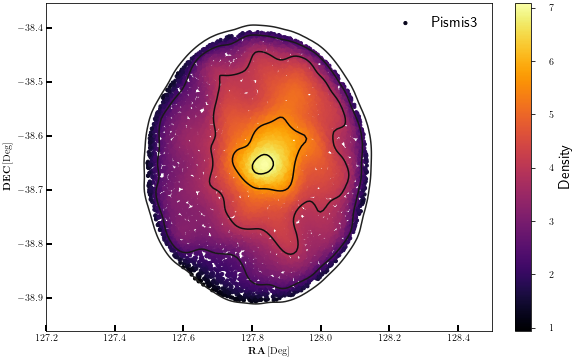}
\caption{ Same as Fig. 2, but for the cluster Pismis3.}
\label{Figure2E}
\end{center}
\end{figure*}

\begin{figure*}[ht] % Changed to [ht] for more flexibility in placement
\centering
\includegraphics[height=10cm]{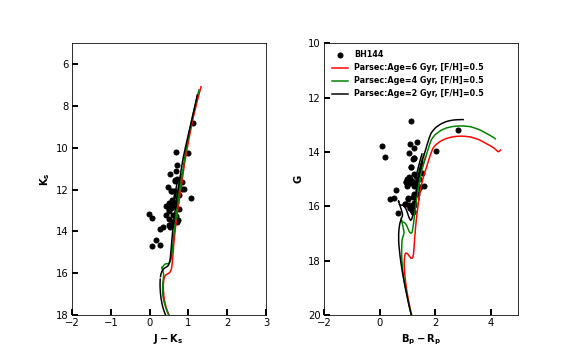}\\
\caption{Colour-magnitude diagrams fo BH144 cluster studied in both NIR and optical  \textit{Gaia} bands. These CMDs are fitted with Parsec isochrones to infer the age and [Fe/H] for each cluster. }
\label{Figure5B}
\end{figure*}

% \begin{figure}[ht] % 
% \centering
% \includegraphics[height=10cm]{SchusterCMDFinal.png}
% \caption{Same as Figure 11 but for Schuster. For clusters with high optical extinction A$_G$ $>$4 , CMDs are not presented.}
% \label{Figure5C}
% \end{figure}

\begin{figure*}[ht] % Changed to [ht] for more flexibility in placement
\centering
\includegraphics[height=8.5cm]{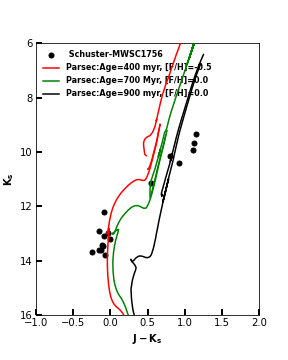}
%\caption{Same as Figure 11 but for Schuster. For clusters with high optical extinction A$_G$ $>$4 , CMDs are not presented.}
\includegraphics[height=8.5cm]{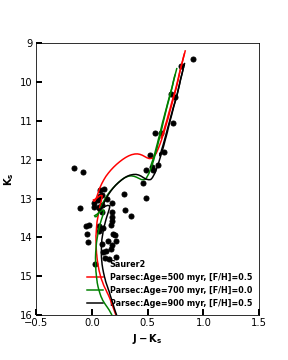}

\includegraphics[height=10cm]{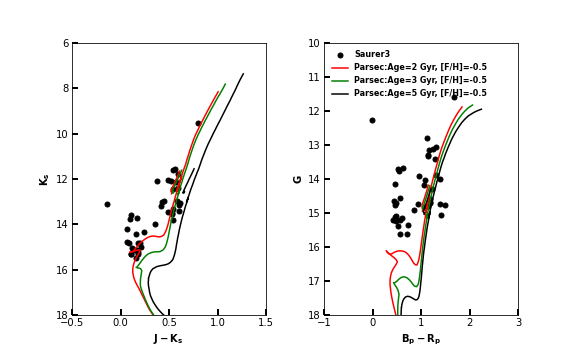}
\caption{Same as Figure 11 but for Schuster, Saurer2 (upper-left and upper-right panels), and Saurer 3 (bottom panel). For clusters with high optical extinction A$_G$ $>$4 , CMDs are not presented.}
\label{Figure5C}
\end{figure*}

\begin{figure*}[ht] % Changed to [ht] for more flexibility in placement
\centering
\includegraphics[height=10cm]{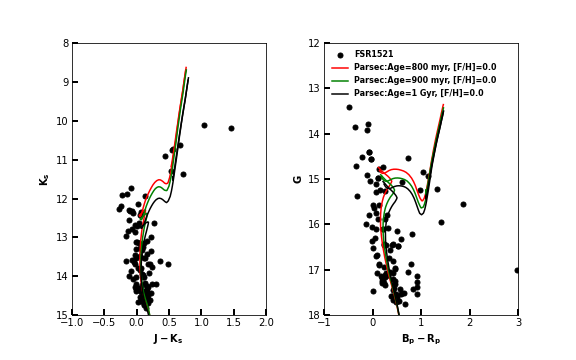}

\includegraphics[height=10cm]{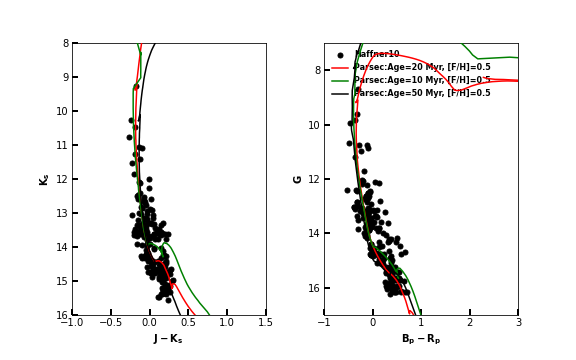}
\caption{Same as Figure 11 but for FSR1521 and Haffner10. For FSR1521, despite the extinction value A$_V$>4, the Gaia CMD is well populated in the main sequence, distinguishing it from other CMDs with similar extinction values. Consequently, we calculated the distance modulus using the distance measured from parallax.}
\label{Figure5D}
\end{figure*}

\begin{figure*}[ht] % Changed to [ht] for more flexibility in placement
\centering
\includegraphics[height=10cm]{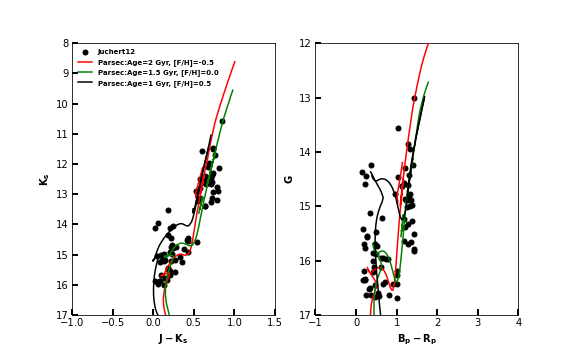}

\includegraphics[height=10cm]{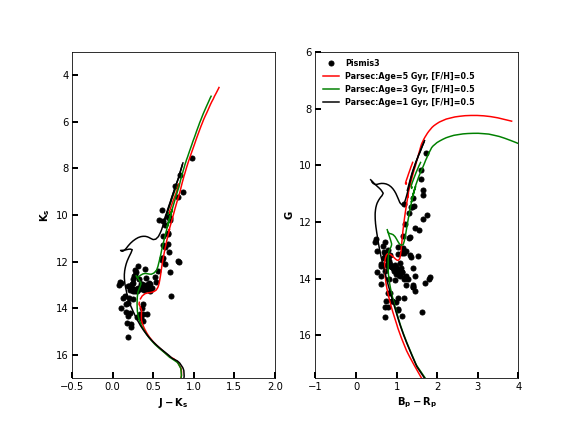}
\caption{Same as Figure 11 but for Juchert12 and Pismis3.}
\label{Figure5E}
\end{figure*}

\begin{figure*}[ht] % Changed to [ht] for more flexibility in placement
\centering
\includegraphics[height=8cm]{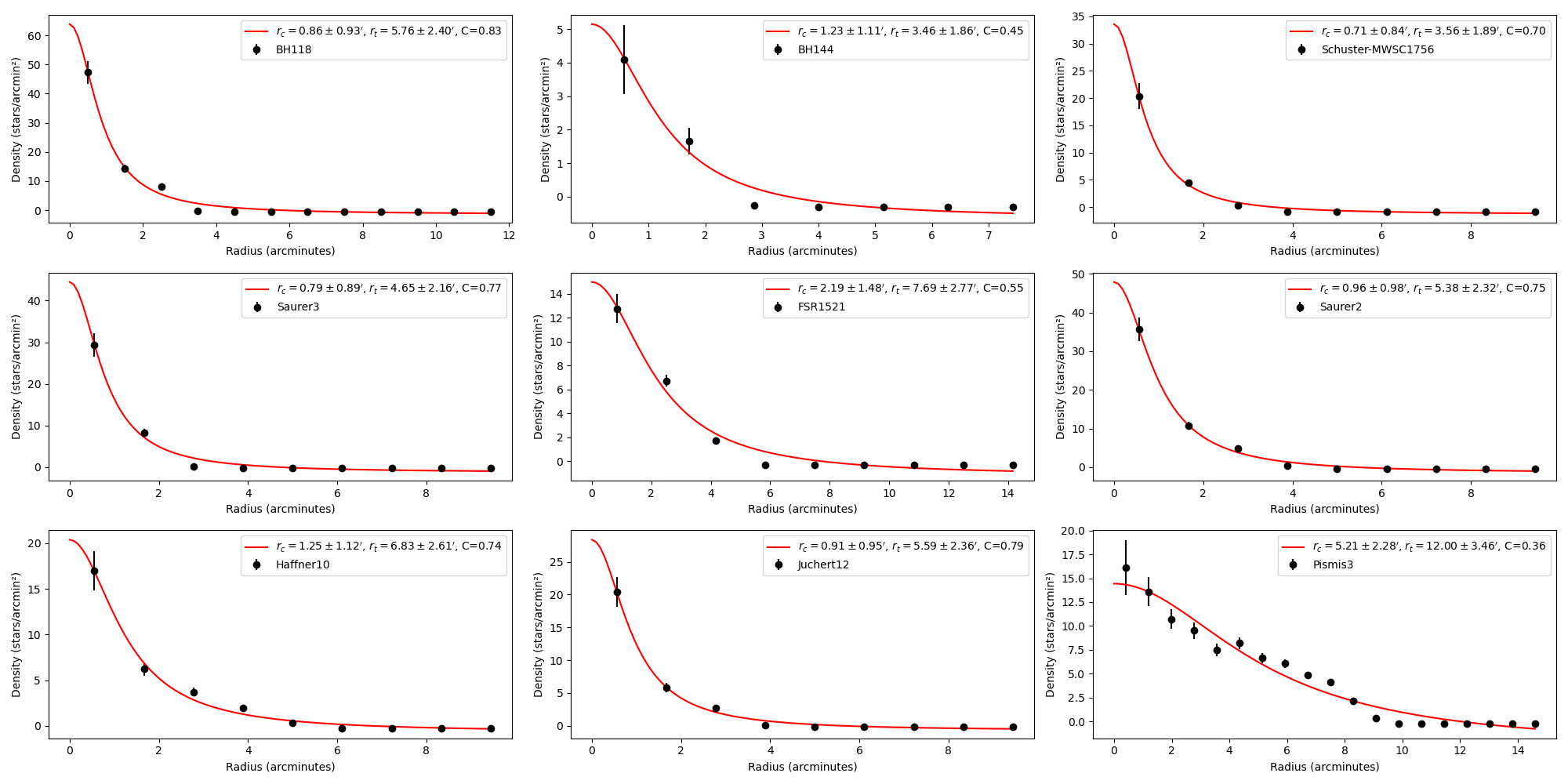}
\caption{ The radial density profile of the clusters is presented, with black points representing the number of stars per unit area within an annular bin. The red line indicates the best-fit King model profile \citep{king1962structure}.}
\label{Figure7}
\end{figure*}

% \begin{figure*}[ht] % Changed to [ht] for more flexibility in placement
% \centering
% \includegraphics[height=5cm]{fsr1521_stru.png}
% \includegraphics[height=5cm]{Saurer2_stru.png}\\
% \includegraphics[height=5cm]{Haffner10_stru.png}
% \includegraphics[height=5cm]{Juchert12_stru.png}
% \includegraphics[height=5cm]{Pismis3_stru.png}

% \caption{ Continuation of figure 16.}
% \label{Figure7B}
% \end{figure*}

\end{appendix}
\end{document}